\documentclass[usenatbib]{mn2e} 
\usepackage{graphicx,fixltx2e,amsmath,amssymb,amsfonts,latexsym,wasysym}

\def\chisqr{\hbox{$\chi^2_{\rm r}$}}
\def\msun{\hbox{${\rm M}_{\odot}$}}

\def\mspy{\hbox{${\rm M}_{\odot}$\,yr$^{-1}$}}
\def\rsun{\hbox{${\rm R}_{\odot}$}}
\def\lsun{\hbox{${\rm L}_{\odot}$}}

\def\rstar{\hbox{$R_{\star}$}}

\def\teff{\hbox{$T_{\rm eff}$}}
\def\logg{\hbox{$\log g$}}
\def\sn{\hbox{S/N}}

\def\kms{\hbox{km\,s$^{-1}$}}
\def\vsini{\hbox{$v \sin i$}}

\def\ptt{\hbox{$10^{-4} I_{\rm c}$}}

\def\degr{\hbox{$^\circ$}}

\newcommand{\caii}{\hbox{Ca$\;${\sc ii}}}
\newcommand{\fei}{\hbox{Fe$\;${\sc i}}}

\newcommand{\hei}{\hbox{He$\;${\sc i}}}
\newcommand{\hal}{\hbox{H${\alpha}$}}
\newcommand{\hbe}{\hbox{H${\beta}$}}

\begin{document}

\title[Magnetism \& accretion of V4046~Sgr]{The close classical T~Tauri binary V4046~Sgr: 
Complex magnetic fields \& distributed mass accretion.  }  
\makeatletter

\def\newauthor{%
  \end{author@tabular}\par
  \begin{author@tabular}[t]{@{}l@{}}}
\makeatother
 
\author[J.-F.~Donati et al.]
{\vspace{1.7mm}
J.-F.~Donati$^1$\thanks{E-mail: 
donati@ast.obs-mip.fr }, 
S.G.~Gregory$^2$, T.~Montmerle$^{3,4}$, A.~Maggio$^5$, C.~Argiroffi$^{6,5}$, \\ 
\vspace{1.7mm}
{\hspace{-1.5mm}\LARGE\rm 
G.~Sacco$^7$, G.~Hussain$^8$, J.~Kastner$^7$, S.H.P.~Alencar$^9$, M.~Audard$^{10}$, J.~Bouvier$^4$, }  \\ 
\vspace{1.7mm}
{\hspace{-1.5mm}\LARGE\rm
F. ~Damiani$^5$, M.~G\"udel$^{11}$, D.~Huenemoerder$^{12}$, G.A.~Wade$^{13}$} \\
$^1$ IRAP--UMR 5277, CNRS \& Univ.\ de Toulouse, 14 Av.\ E.~Belin, F--31400 Toulouse, France \\
$^2$ California Institute of Technology, MC 249-17, Pasadena, CA 91125, USA \\ 
$^3$ Institut d'Astrophysique de Paris, 98bis Bd Arago, F--75014 Paris, France \\ 
$^4$ IPAG--UMR 5274, CNRS \& Univ.\ J.~Fourier, 414 rue de la Piscine, F--38041 Grenoble, France \\ 
$^5$ INAF, Osservatorio Astronomico di Palermo, Piazza del Parlamento 1, 90134 Palermo, Italy \\ 
$^6$ Dip.\ di Fisica, Univ.\ di Palermo, Piazza del Parlamento 1, 90134 Palermo, Italy \\
$^7$ Center for Imaging Science, Rochester Institute of Technology, 54 Lomb Memorial Drive, Rochester, NY 14623, USA \\ 
$^8$ ESO, Karl-Schwarzschild-Str.\ 2, D-85748 Garching, Germany \\ 
$^9$ Departamento de F\`{\i}sica -- ICEx -- UFMG, Av. Ant\^onio Carlos, 6627, 30270-901 Belo Horizonte, MG, Brazil \\ 
$^{10}$ ISDC Data Center for Astrophysics, University of Geneva, Ch. d'Ecogia 16, 1290 Versoix, Switzerland \\ 
$^{11}$ Department of Astronomy, University of Vienna, Türkenschanzstr.~17, 1180 Vienna, Austria\\ 
$^{12}$ Massachusetts Institute of Technology, Kavli Institute for Astrophysics and Space Research, 70 Vassar St., Cambridge, MA 02139, USA \\ 
$^{13}$ Department of Physics, Royal Military College of Canada, PO Box 17000, Station Forces, Kingston, Ontario K7K~7B4, Canada
}

\date{2011 April, MNRAS submitted}
\maketitle
 
\begin{abstract}  

We report here the first results of a multi-wavelength campaign focussing on magnetospheric accretion 
processes within the close binary system V4046~Sgr, hosting two partly-convective classical T~Tauri 
stars of masses $\simeq$0.9~\msun\ and age $\simeq$12~Myr.  
In this paper, we present time-resolved spectropolarimetric observations collected in 2009 September 
with ESPaDOnS at the Canada-France-Hawaii Telescope (CFHT) and covering a full span of 7~d or 
$\simeq$2.5 orbital/rotational cycles of V4046~Sgr.  
Small circularly polarised Zeeman signatures are detected in the photospheric absorption lines but 
not in the accretion-powered emission lines of V4046~Sgr, thereby demonstrating that both system 
components host large-scale magnetic fields weaker and more complex than those of younger, fully-convective 
cTTSs of only a few Myr and similar masses.  

Applying our tomographic imaging tools to the collected data set, we reconstruct maps of the large-scale 
magnetic field, photospheric brightness and accretion-powered emission at the surfaces of both stars of 
V4046~Sgr.  We find that these fields include significant toroidal components, and that their poloidal 
components are mostly non-axisymmetric with a dipolar component of 50-100~G strongly tilted with respect 
to the rotation axis;  given the similarity with fields of partly-convective main-sequence stars 
of similar masses and rotation periods, we conclude that these fields are most likely generated by 
dynamo processes.  We also find that both stars in the system show cool spots close to the pole and
extended regions of low-contrast, accretion-powered emission;  it suggests that mass accretion is 
likely distributed rather than confined in well defined high-contrast accretion spots, in agreement 
with the derived magnetic field complexity.  
\end{abstract}


\section{Introduction} 
\label{sec:int}

From the collapse of giant molecular clouds to their fragmentation into individual 
stars and their planetary systems, from the dissipation of the initial cloud angular 
momentum content to the generation of outflows and jets, magnetic fields have a 
strong impact on most physical processes involved in the formation of stars and 
planets and are now clearly recognized as one of the few main ingredients (along 
with gravitation and turbulence) in Nature's recipe to build new worlds 
\citep[e.g.,][]{Andre09, Donati09}.  
In the latter formation phases in particular, low-mass Sun-like stars young enough 
to be still surrounded by, and accreting mass from, a gaseous disc (the classical 
T~Tauri stars or cTTSs) are apparently capable of generating and sustaining a 
large-scale magnetic field through dynamo processes, which in turn manages to disrupt 
the inner disc regions and drastically brake the rotation of the central protostar 
\citep[see, e.g.,][for a review]{Bouvier07}.  

The presence of intense magnetic fields at the surfaces of cTTSs, and more generally 
of all T~Tauri stars, was first derived 
through indirect proxies \citep[i.e., continuum or line emission throughout 
the whole electromagnetic spectrum, from the radio to X-rays, e.g.,][for 
reviews]{Feigelson99, Favata03, Gudel09}, 
then demonstrated directly (through the Zeeman broadening of spectral lines) about 2 
decades ago \citep[e.g.,][for an overview]{Johns07}.  However, the actual 
large-scale magnetic topologies of cTTSs - a crucial parameter to elucidate the way 
such fields manage to couple the protostars to their discs, to funnel the accreted 
disc material to the stellar surface, and to strongly brake the rotation of the 
protostar - remained rather elusive until recently.  Through
spectropolarimetric observations consisting of time-series of Zeeman signatures from 
accreting and non-accreting regions at the surfaces of cTTSs, new studies demonstrated 
that very young low-mass stars indeed host large-scale fields of dynamo origin, whose 
topologies are reminiscent of those of main-sequence stars (once differences in the 
internal stellar structures are taken into account).  

This new opportunity offers the option of investigating magnetospheric accretion processes 
of cTTSs in a much more global and consistent way, by carrying out simultaneously 
observations of selected prototypical cTTSs with as wide a spectral coverage as possible, 
and including in particular the X-ray and optical domains.  While optical lines provide 
information on the large-scale magnetosphere and on the location and rate at which mass is 
accreted from the disc to the surface of the protostar, high resolution X-ray spectra (and 
in particular the softer component) yields key material on the shock that occurs when the 
accretion flows collide with the high atmosphere of the protostar.  
The present study takes place in the framework of an international multi-wavelength, 
multi-site campaign organised on V4046~Sgr, one of the few known close cTTS binaries 
and one of the best cases where the softer component of the X-ray spectrum (attributable 
to magnetospheric accretion shocks) can be recorded with sufficient accuracy \citep{Gunther06}.  
This campaign was triggered by a Large Program with XMM-Newton, aimed at obtaining 
phase-resolved X-ray observations of V4046~Sgr covering about 2 consecutive 
orbital/rotation cycles of the cTTS binary (for a total exposure time of $\simeq$370~ks).  

Spectropolarimetric (optical) observations of V4046~Sgr were also scheduled in conjunction 
with (though slightly before than) the main X-ray program, and were carried out in a way 
similar to those achieved in the framework of the international MaPP (Magnetic Protostars 
and Planets) project.  Through a survey of a dozen cTTSs, MaPP investigates how the 
large-scale magnetic fields of low-mass protostars depend on stellar parameters such as 
mass, age, rotation and accretion rates \citep[e.g.,][]{Donati10b, Donati11};  MaPP also 
includes a theoretical component aiming at describing consistently (through analytical modelling 
and numerical simulations) how magnetic fields of cTTSs couple to their surrounding accretion 
disc, how they channel accretion into discrete funnels and what is the resulting angular 
momentum evolution \citep[see, e.g.,][for an extensive review on the subject]{Gregory10}.  
At this stage, however, MaPP has only focussed on single (or distant binary) stars and has 
not addressed the specific issue of magnetospheric accretion processes in close binary 
stars;  the present campaign on V4046~Sgr thus appears as a worthwhile and timely complement 
to the core MaPP program.  

The complete results of the V4046~Sgr project will be published in a series of papers, 
starting with a global overview of the campaign and a summary of the main X-ray 
results (Argiroffi et al.\ 2011, submitted).  
The present paper belongs to this series and aims at obtaining, for both cTTSs 
forming the close V4046~Sgr binary, observational constraints similar to those 
recently derived for single cTTSs;  in particular, it aims at simultaneously deriving 
the large-scale magnetic fields of both stars of V4046~Sgr, their photospheric 
brightness maps and their chromospheric distributions of accretion-induced hot spots.  
Companion papers (e.g., Maggio et al.\ 2011, Sacco et al.\ 2011, in 
preparation) describe additional data sets collected during the same campaign.  

After summarising the main characteristics of the close V4046~Sgr cTTS binary 
(Sec.~\ref{sec:v4046}), we describe the new spectropolarimetric observations we 
collected and outline the associated temporal variability and rotational modulation 
(Secs.~\ref{sec:obs} and ~\ref{sec:var}).  We then detail the modelling of these 
data with our magnetic imaging code (Sec.~\ref{sec:mod}) and briefly compare our 
results with those already published for single cTTSs and with the predictions of 
the latest numerical simulations of accretion from circumbinary discs in close 
cTTS binaries (Sec.~\ref{sec:dis}).

\section{V4046~Sgr (HDE~319139, HBC~662, AS~292) }
\label{sec:v4046}

Located at a distance of about 73~pc, V4046~Sgr is a likely member of the nearby 
loose $\beta$~Pic association \citep{Torres08}.  It is one of the few known close 
cTTS binaries, showing signs of accretion typical of single cTTSs (e.g., large and 
variable \hal\ emission).  It is a double line spectroscopic binary with a circular orbit, 
a rather short orbital period ($\simeq$2.42~d) and a prominent excess at IR and radio 
frequencies indicating the presence of a circumbinary accretion disc 
\citep[with an inner radius of 0.2--0.4~AU and extending out to several hundred 
AUs,][]{Quast00, Rodriguez10}.  The photometric and spectroscopic periods are 
equal and demonstrate that the system is synchronised, as expected for a binary system 
as close as V4046~Sgr.  Given that V4046~Sgr has already completed both circularisation and 
synchronisation, we can logically assume that the orbital and rotation axes are aligned, 
following predictions of the tidal theory \citep[e.g.,][]{Hut81, Zahn89}.   

Two different detailed spectroscopic studies of V4046~Sgr are available from the 
literature \citep{Quast00, Stempels04}.  From the photometric colors, velocity 
amplitudes and the wavelength-dependent primary-to-secondary magnitude contrast, these 
studies show in particular that both system components have similar masses and surface 
temperatures of about 0.9~\msun\ and 4250~K, the primary being about 6\% more massive 
and 250~K hotter than the secondary.  
The system is seen mostly pole-on, with an angle between the orbital axis and the 
line of sight of $\simeq$35\degr\ \citep{Quast00, Rodriguez10}.  
The line-of-sight-projected equatorial rotation velocities \vsini\ that we estimate 
for both system components (equal to $13.5\pm0.5$ and $12.5\pm0.5$~\kms\ for the 
primary and secondary stars respectively) agree with previous estimates and indicate
respective radii of $1.12\pm0.05$ and $1.04\pm0.05$~\rsun.  
The position of both stars in the HR diagram (as derived from the above temperatures and 
radii) and their relative locations with respect 
to the theoretical evolutionary tracks and isochrones of \citet{Siess00}, shown in 
Fig.~\ref{fig:hrd}, suggest an age of $\simeq$15~Myr;  this is in agreement with recent 
age estimates of the $\beta$~Pic moving group as a whole \citep[ranging from 10 to 20~Myr, 
e.g.,][]{Mentuch08, daSilva09, Yee10} and in particular with the widely accepted 
and quoted age of 12~Myr \citep{Torres08}.  
It indicates that V4046~Sgr is approaching the end of the TTS stage;  
at this age, both stars should be partly convective, with the convective zone occupying 
the outer $\simeq$50\% of the stellar radius \citep{Siess00}.  

\begin{figure}
\includegraphics[scale=0.35,angle=-90]{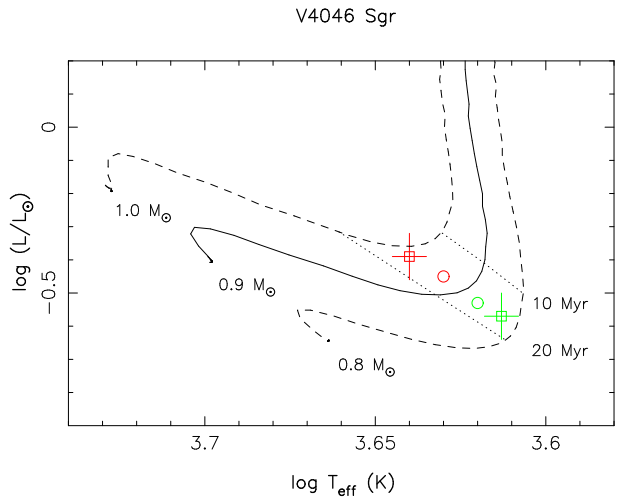}
\caption[]{Observed (open squares and error bars) location 
of the primary (red) and secondary (green) components of V4046~Sgr in the HR diagram, 
as derived from their respective temperatures and radii (see text).  
The best match to these positions when further imposing a primary-to-secondary mass 
ratio of 1.06 (as derived from the velocity curves) is also indicated (open circles).  
The PMS evolutionary tracks and corresponding isochrones \citep{Siess00} assume solar 
metallicity and include convective overshooting.  } 
\label{fig:hrd}
\end{figure}

The distance between both components as derived from the semi-amplitude of the 
velocity curves (found to be $51.4\pm0.2$ and $54.3\pm0.2$~\kms\ from our data) is 
8.8~\rsun\ or 0.041~AU, in good agreement with the results of \citet{Quast00}.  
However, the system velocity that we derive ($-5.7\pm0.2$~\kms, see 
Sec.~\ref{sec:mod}) is significantly different from the former estimate \citep[e.g., 
$-6.94\pm0.16$~\kms][]{Quast00} and suggests that the system also hosts a third distant 
component.  The very precise radial velocity (RV) of the circumbinary (CO disc) material 
\citep[equal to $-6.21\pm0.01$~\kms\ in the heliocentric frame,][]{Rodriguez10} 
confirms that this is likely the case.  

X-ray spectra of V4046~Sgr with Chandra-HETGS clearly show He-like triplets with 
line strength ratios in agreement with the predictions of magnetospheric accretion 
models, confirming that V4046~Sgr is still accreting mass from the surrounding disc 
\citep{Gunther06}.  From the equivalent widths and the corresponding line fluxes of 
the optical emission lines usually considered as good accretion proxies 
(and in particular the \hei\ $D_3$ line and the \caii\ infrared triplet, IRT) and using 
the published empirical correlations between lines and accretion fluxes \citep{Fang09}, 
we can derive an estimate of the average logarithmic mass accretion rate (in \mspy) 
at the surface of each star of V4046~Sgr, that we find to be equal to $-9.3\pm0.3$ 
(assuming both stars have equal accretion rates) in agreement with independent estimates 
from optical proxies \citep{Curran11}.  As usual, this is larger than the estimate derived 
from X-ray data \citep[equal to $-9.84\pm0.11$ for V4046~Sgr,][]{Curran11}.  Optical 
veiling, i.e., the apparent weakening of the photospheric spectrum (presumably caused by 
accretion) is apparently weak for V4046~Sgr \citep[e.g.,][]{Stempels04}.  

V4046~Sgr shares similarities with another member of the $\beta$~Pic 
moving group, the young pre-main-sequence close binary HD~155555 \citep{Dunstone08a, 
Yee10}, whose stars are slightly more massive (1.1--1.2~\msun), 
larger ($\simeq$1.3~\rsun), warmer ($\simeq$5500~K) and closer to each other (0.036~AU);  
both systems are thus structurally alike though different regarding their accretion 
properties and related issues, HD~155555 having dissipated its accretion disc already.  
HD~155555 may thus serve as a convenient check to identify what in V4046~Sgr is typical 
of young partly-convective Sun-like binaries and what is more specific to cTTS binaries.

\section{Observations}
\label{sec:obs}

Spectropolarimetric observations of V4046~Sgr were collected from 2009 September 03 to 09 
using ESPaDOnS on the CFHT.  ESPaDOnS collects stellar spectra spanning the whole optical domain
(from 370 to 1,000~nm) at a resolving power of 65,000 (i.e., 4.6~\kms) and with a spectral
sampling of 2.6~\kms, in either circular or linear polarisation \citep{Donati03}.
A total of 8 circular polarisation spectra were collected over a timespan of 7 consecutive nights.  
All polarisation spectra consist of 4 individual subexposures each lasting
390~s and taken in different polarimeter configurations to allow the removal of
all spurious polarisation signatures at first order.

All raw frames are processed with {\sc Libre~ESpRIT}, a fully automatic reduction
package/pipeline available at CFHT.  It automatically performs optimal
extraction of ESPaDOnS unpolarized (Stokes $I$) and circularly polarized (Stokes $V$)
spectra grossly following the procedure described in \citet{Donati97b}.
The velocity step corresponding to CCD pixels is about 2.6~\kms;  however, thanks
to the fact that the spectrograph slit is tilted with respect to the CCD lines,
spectra corresponding to different CCD columns across each order feature a
different pixel sampling.  {\sc Libre~ESpRIT} uses this opportunity to carry out
optimal extraction of each spectrum on a sampling grid denser than the original
CCD sampling, with a spectral velocity step set to about 0.7 CCD pixel
(i.e.\ 1.8~\kms).
All spectra are automatically corrected of spectral shifts resulting from
instrumental effects (e.g., mechanical flexures, temperature or pressure variations)
using telluric lines as a reference.  Though not perfect, this procedure provides
spectra with a relative RV precision of better than 0.030~\kms\
\citep[e.g.,][]{Donati08b}.

\begin{table}
\caption[]{Journal of observations collected in 2009 September.   
Columns $1-4$ respectively list the UT date, the heliocentric Julian date and 
UT time (both at mid-exposure), and the peak signal to noise ratio (per 2.6~\kms\ 
velocity bin) of each observation (i.e., each sequence of $4\times390$~s subexposures).  
Column 5 lists the rms noise level (relative to the unpolarized continuum level 
$I_{\rm c}$ and per 1.8~\kms\ velocity bin) in the circular polarization profile 
produced by Least-Squares Deconvolution (LSD), while column~6 indicates the 
orbital/rotational cycle associated with each exposure (using the ephemeris given by 
Eq.~\ref{eq:eph}).  }   
\begin{tabular}{cccccc}
\hline
Date & HJD          & UT      &  \sn\  & $\sigma_{\rm LSD}$ & Cycle \\
     & (2,455,000+) & (h:m:s) &      &   (\ptt)  & (3336+) \\
\hline
Sep 03 & 77.77383 & 06:33:30 & 180 & 2.9 & 0.755 \\
Sep 04 & 78.81648 & 07:35:02 & 170 & 2.6 & 1.186 \\
Sep 05 & 79.72103 & 05:17:41 & 200 & 2.1 & 1.560 \\
Sep 06 & 80.72065 & 05:17:14 & 200 & 2.1 & 1.972 \\
Sep 06 & 80.81321 & 07:30:32 & 210 & 1.9 & 2.011 \\
Sep 07 & 81.72152 & 05:18:36 & 160 & 2.5 & 2.386 \\
Sep 08 & 82.72171 & 05:18:60 & 200 & 2.0 & 2.799 \\
Sep 09 & 83.72107 & 05:18:11 & 210 & 1.8 & 3.212 \\
\hline
\end{tabular}
\label{tab:logesp}
\end{table}

The peak signal-to-noise ratios (\sn, per 2.6~\kms\ velocity bin) achieved on the
collected spectra (i.e., the sequence of 4 subexposures) range between 160 and
210 depending on weather/seeing conditions, with a median of 200.  

Following \citet{Stempels04}, orbital/rotational cycles $E$ are computed from 
heliocentric Julian dates according to the ephemeris:  
\begin{equation}
\mbox{HJD} = 2446998.335 + 2.4213459 E 
\label{eq:eph}
\end{equation}
We however note that the time of first conjunction (with the primary component 
in front) in our data is significantly shifted with respect to the ephemeris predictions 
and occurs at phase 0.681\footnote{The epoch of first quadrature as derived from our 
observations occurs on HJD~=~2455078.199$\pm$0.001~d, with one of our spectra collected 
very shortly after second conjunction (on Sep~04).} instead of phase 0.75 (i.e., 
0.069~cycle or about 4~hr ahead of time).  
The origin of this shift is not fully clear yet.  It can reflect a slightly 
overestimated orbital period;  if this is the case, the orbital period needs to be updated 
to 2.421296$\pm$0.000001~d, i.e., a value smaller than those of \citet{Quast00} and of 
\citet{Stempels04} by 9 and 12 of their $\sigma$ (0.000004~d) respectively.  It may also potentially 
reflect (at least partly) temporal fluctuations of the orbital period, either caused by a 
light-time effect (e.g., that induced by the distant third body causing the RV changes 
mentioned in Sec.~\ref{sec:v4046}) and/or by changes in the quadrupolar moments of both 
components of V4046~Sgr \citep[resulting from activity cycles of one or both system 
components and known as the Applegate effect,][]{Applegate92, Lanza06}.  This issue will 
be specifically addressed in more details in a forthcoming paper.  

\begin{figure}
\includegraphics[scale=0.35,angle=-90]{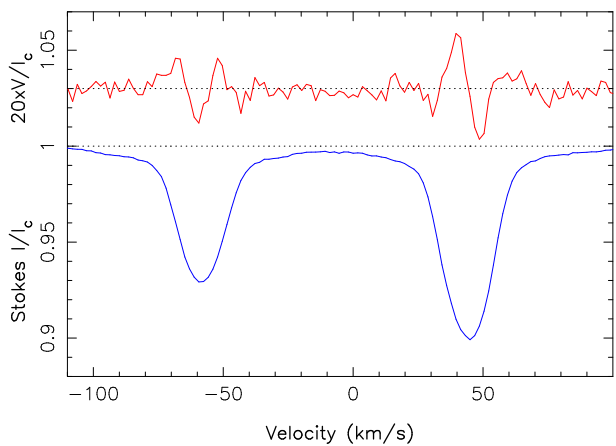}
\caption[]{LSD circularly-polarized (Stokes $V$) and unpolarized (Stokes $I$)
profiles of V4046~Sgr (top/red, bottom/blue curves respectively) collected on 2009 
September 06 (cycle 1.972), i.e., slightly after first quadrature (occurring at 
phase 0.931, see Sec.~\ref{sec:obs}).  Zeeman signatures are clearly detected 
in the LSD profiles of both the primary (right) and secondary (left) system components.
The mean polarization profile is expanded by a factor of 20 and shifted upwards
by 1.03 for display purposes.  }
\label{fig:lsd}
\end{figure}

Least-Squares Deconvolution \citep[LSD,][]{Donati97b} was applied to all
observations.   The line list we employed for LSD is computed from an {\sc
Atlas9} LTE model atmosphere \citep{Kurucz93} and corresponds to a K5V 
spectral type ($\teff=4,250$~K and  $\logg=4.5$) appropriate for V4046~Sgr.
Only moderate to strong atomic spectral lines (with line-to-continuum core 
depressions larger than 40\% prior to all non-thermal broadening) are included 
in this list;  spectral regions with strong lines mostly formed outside the 
photosphere (e.g., Balmer, He, \caii\ H, K and IRT lines) and/or heavily crowded 
with telluric lines were discarded.  
Altogether, about 8,000 spectral features (with about 40\% from \fei) are used 
in this process.  
Expressed in units of the unpolarized continuum level $I_{\rm c}$, the average 
noise levels of the resulting Stokes $V$ LSD signatures are ranging from 1.8 to 
2.9$\times10^{-4}$ per 1.8~\kms\ velocity bin.  

The full journal of observations is presented in Table~\ref{tab:logesp}.  

\begin{figure*}
\center{\includegraphics[scale=0.96,angle=-90]{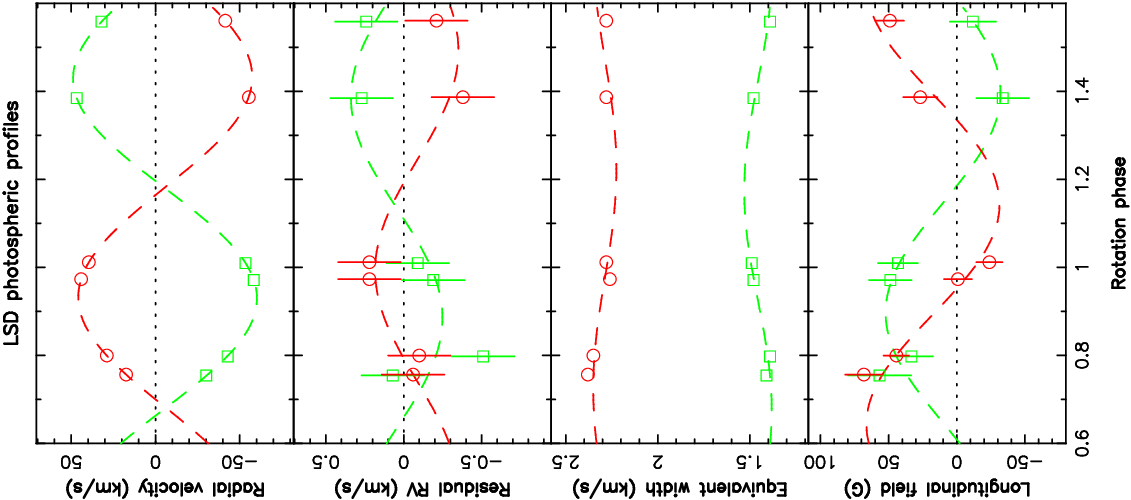}\hspace{4mm}
\includegraphics[scale=0.96,angle=-90]{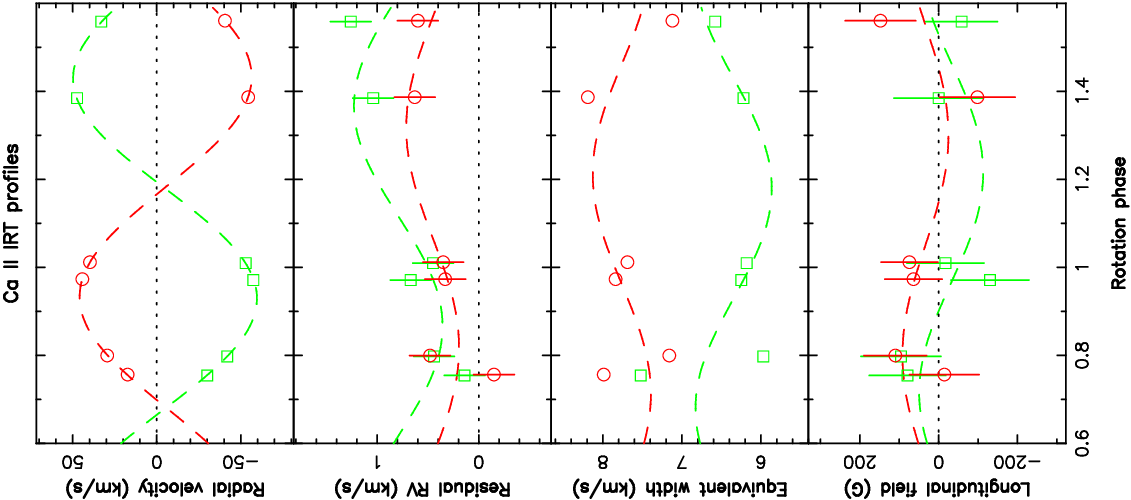}}
\caption[]{Rotational modulation of the RV (top row), residual RV (second row), 
equivalent width (third row) and longitudinal field (bottom row) derived from LSD 
photospheric profiles (left panels) and the average \caii\ IRT line (right panels) 
and for both the primary (red circles) and secondary (green squares) system components 
of V4046~Sgr.  First and second orbital conjunctions respectively occur at phases 0.681 and 0.181 
in the ephemeris used here (see Eq.~\ref{eq:eph}).  Data collected at near-conjunction epochs and 
at which spectral contributions of both components overlap and cannot be easily separated from 
one another (i.e., cycles 1.186 and 3.212) were excluded from this plot.  Fits with sine/cosine waves are 
included (and shown as dashed lines) to outline the amount of variability attributable 
to rotational modulation (whenever significant), but are not used in the subsequent imaging process.  
$\pm1$~$\sigma$ 
error bars on data points, based on photon noise only, are also shown whenever larger than symbols.  }
\label{fig:var}
\end{figure*}

\section{Spectroscopic variability}
\label{sec:var}

Before applying our tomographic imaging tools to the V4046~Sgr data, it is usually worthwhile to 
examine how spectral lines (and in particular the equivalent widths, RVs and the average 
magnetic fluxes) vary with rotation cycle throughout the observing run;  at the very least, it 
provides a rough, intuitive understanding of how the large-scale field is structured and 
oriented, and where the cool photospheric and hot accretion spots are likely to be 
located - to be compared later on with the detailed results of the full imaging process 
(described in the next section).  This is what we present in this section;  note that 
the 2 spectra recorded near second conjunction (at cycles 1.186 and 3.212) were excluded 
from part of this preliminary analysis, since the contributions of both system components 
cannot be easily separated from one another.  

The main modulation observed in LSD profiles of photospheric lines of V4046~Sgr is 
the RV shifts resulting from the orbital motion of both system components;  
Fig.~\ref{fig:lsd} shows an example profile recorded at cycle 1.972, i.e., slightly 
after the first quadrature (occurring at phase 0.931, see Sec.~\ref{sec:obs}) and 
when both components 
are clearly separated from one another, while Fig.~\ref{fig:var} (top left panel)  
shows the phased RV curves of both components.  The orbital elements derived from 
these profiles (as a by-product of the imaging procedure described in 
Sec.~\ref{sec:mod}) agree with those of \citet{Quast00} (except for the significant 
shift in the phase of first conjunction, see Sec.~\ref{sec:obs}, and for the change in 
the system velocity, see Sec.~\ref{sec:v4046}) but do not support 
the newer estimates of \citet{Stempels04}.  

Zeeman signatures are detected at all times in Stokes $V$ LSD profiles and in 
close association with the spectral lines of both components (see Fig.~\ref{fig:lsd} 
for an example).  In most spectra, Stokes $V$ signatures are complex, with a typical 
peak-to-peak amplitudes of 0.2\%;  they feature several reversals throughout the line 
profile, suggesting that the parent field topology is not simple.  
The line-of-sight projected component of the field averaged over the 
visible stellar hemisphere and weighted by brightness inhomogeneities 
\citep[called the longitudinal field and estimated from the first 
moment of the Stokes $V$ profile, e.g.,][]{Donati97b} is 
rather weak (always smaller than 70~G and most of the time smaller than 50~G, with 
typical error bars of $\simeq$10~G) 
and poorly informative about the large-scale field (see Fig.~\ref{fig:var} 
lower left panel).   

LSD Stokes $I$ profiles also show a small level of rotational modulation (in addition 
to the RV shifts caused by orbital motion\footnote{The RV shifts of spectral lines 
resulting from the orbital motion of both system components were estimated as part 
of the imaging process (see Sec.~\ref{sec:mod}) and are thus free of contamination 
from cool surface spots \citep[as opposed to those of][]{Quast00};  they are thus 
the most logical ones to use when looking for potential RV signatures of cool 
surface spots.}) as a likely consequence of the presence 
of cool spots on the stellar surface of both system components (also causing the 
reported photometric modulation).  The line equivalent widths of both system components 
are observed to vary in anticorrelation and directly reflect the fluctuations in the 
secondary to primary brightness ratio (around a mean of about 0.64, see Fig.~\ref{fig:var} 
third left panel);  the primary star is brightest and/or the secondary star is faintest 
by a few \% around the first orbital conjunction (i.e., phase 0.681) while the opposite holds around 
the second orbital conjunction (i.e., phase 0.181).  Once corrected from the spectral dilution 
of the companion, the LSD profiles of both system components show equivalent widths of about 
3.9~\kms, fully compatible with those of non-accreting young stars of similar spectral 
types (e.g., V410~Tau);  this demonstrates that veiling was weak (i.e., $<$5\%) for V4046~Sgr 
at the time of our observations.  

The residual RVs of both system components (in the rest frame of each star, i.e., with 
the orbital motion removed) is also showing a low-amplitude modulation of $<1$~\kms\ 
peak to peak, see Fig.~\ref{fig:var} second left panel).  As the \vsini 's are much larger
than the amplitude of this modulation, this suggests in particular the 
presence of high-latitude spots, around phase 0.2 on the primary star and phase 0.7 on 
the secondary star, in full agreement with the conclusions derived from the rotational 
modulation in the equivalent widths of LSD profiles of both components (see above).  

We note that LSD Stokes $I$ and $V$ profiles of both system components repeat well 
from one rotation cycle to the next, further supporting the conclusion that rotational 
modulation largely dominates over intrinsic variability in V4046~Sgr, even more than in 
the moderately accreting cTTS V2129~Oph \citep{Donati11}.  

\begin{figure}
\includegraphics[scale=0.35,angle=-90]{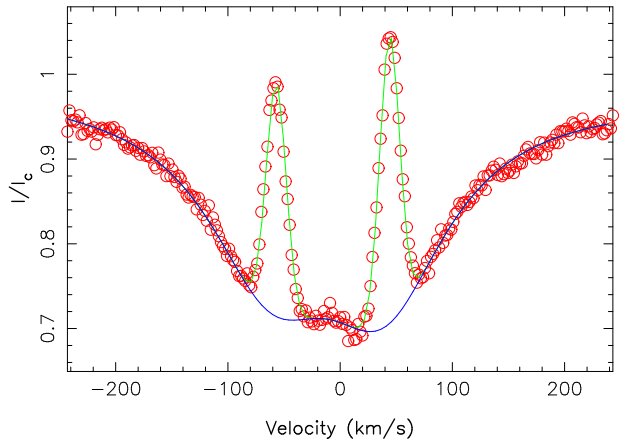}
\caption[]{Double Lorentzian fit (bottom blue line) with additional double Gaussian fit (top 
green line) to the observed mean IRT line profile of V4046~Sgr (red circles) on 2009 Sep 06 (cycle 1.972).  
The bottom blue line is the background profile model that we subtract from the observed profile to 
recover the emission profiles of both system components. }
\label{fig:irt}
\end{figure}

As in previous papers, we use core emission in \caii\ IRT lines as a proxy of surface 
accretion even though it usually features a significant (and often dominant) 
contribution from the non-accreting chromosphere.  Previous work \citep{Donati10b, Donati11} 
demonstrated in particular that rotational modulation in Stokes $I$ and $V$ \caii\ IRT 
profiles correlates well with that derived from the more conventional \hei\ $D_3$ accretion 
proxy, even for mass accretion rates as low as (or even lower than) that of V4046~Sgr (see below).  
In the particular case of V4046~Sgr, the emission 
contributions of both system components are narrow enough to be well separated in velocity 
during most of the orbital cycle, offering the option of characterizing the accretion 
behaviour of both stars independently from one another.  
In practice, we start by constructing a LSD-like weighted average of the 3 IRT lines;  
we then subtract the underlying (much wider) photospheric absorption profiles, 
with a single Lorentzian fit to the far line wings when the emission peaks of both 
system components overlap (i.e., at cycle 1.186 and 3.212, see Sec.~\ref{sec:mod}), 
and with a double Lorentzian fit to the far wings and the line center when the emission 
peaks of both system components are well separated in velocity (see Fig.~\ref{fig:irt} for 
an example).  

The RVs of the emission peaks are fully compatible with the orbital solution derived 
from photospheric lines (see Fig.~\ref{fig:var} top right panel).  In the stellar 
rest frame of each system component, both emission peaks are slightly shifted redwards 
by about 0.7~\kms\ (see Fig.~\ref{fig:var} second right panel), as usual in cTTSs and 
confirming that it traces slowly-moving regions of the post-shock accretion funnels 
close to the surface of the star.  The equivalent widths of both emission peaks are 
roughly constant with phase and equal to $\simeq$15~\kms\ or 0.040~nm (once corrected 
from the spectral dilution) and vary by less than 2~\kms\ over 
the rotation cycle (see Fig.~\ref{fig:var} third right panel).  It suggests a rather 
low level of accretion and accretion regions either distributed over the whole visible 
hemisphere or located very close to the pole (hence 
producing a very small level of modulation);  this is further evidenced by the 
low-amplitude RV modulation of the emission peaks of both components (once the orbital 
motion is removed, see Fig.~\ref{fig:var} second right panel).  

We detect no Zeeman signatures in conjunction with \caii\ IRT emission, with error bars 
on the corresponding longitudinal fields most of the time smaller than 100~G (see 
Fig.~\ref{fig:var} bottom right panel).  This is fairly unusual in cTTSs, even in those 
with low mass accretion rates \citep[e.g.,][]{Donati11} and suggests that the large-scale 
fields of both system stars have rather weak dipole components.  We nevertheless use 
the corresponding Stokes $V$ profiles as an additional constraint for the magnetic modelling
(see Sec.~\ref{sec:mod}).  

Although usually considered as the most reliable accretion proxy, emission in the \hei\ $D_3$ 
line at 587.562~nm (not shown here) is hardly usable in the particular case of V4046~Sgr.  This 
is mostly the consequence of the relative weakness of this line in the spectrum of V4046~Sgr and 
of its significant width (with the contribution of both components overlapping most of the time) - 
all of this occurring within a relatively crowded spectral region.  The best we can obtain 
is that the total line equivalent width (for both system components as a whole), equal to 
17.5~\kms\ or 0.034~nm in average, is roughly constant with phase throughout the orbital cycle 
(in agreement with what the \caii\ IRT lines show) except for a (presumably sporadic) emission 
burst detected at cycle 2.386 (also detected in Balmer lines, see below, but not in \caii\ IRT 
lines).  No Zeeman signatures are detected 
in association with the \hei\ $D_3$, consistent with the non-detection in \caii\ IRT lines.  

\begin{figure*}
\center{
\includegraphics[scale=0.58,angle=-90]{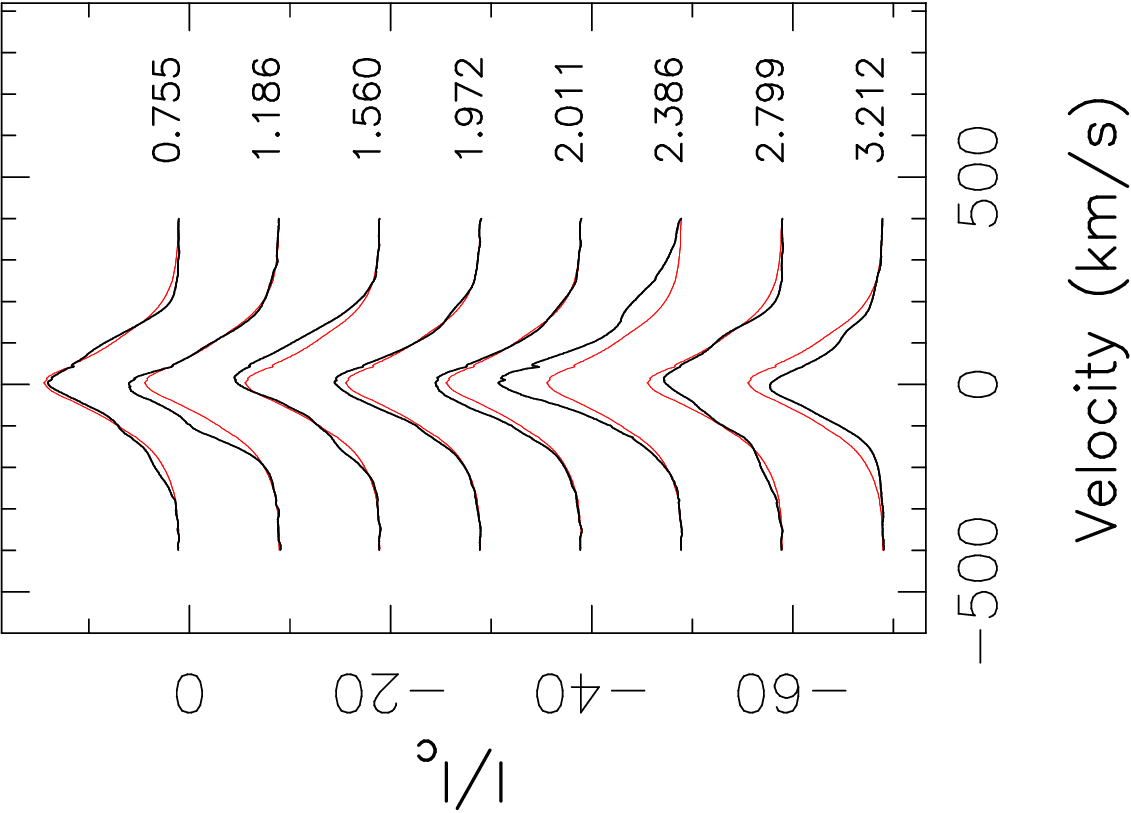}\hspace{5mm}
\includegraphics[scale=0.58,angle=-90]{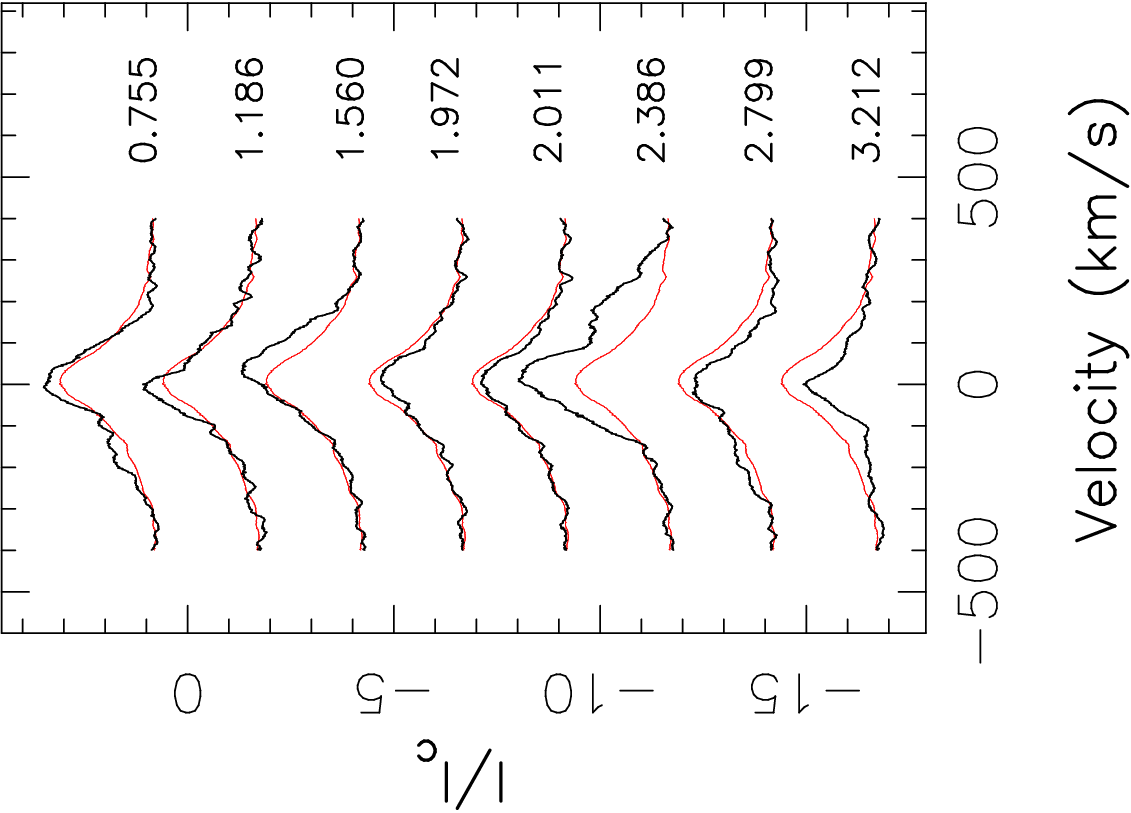}}
\caption[]{Variation of the \hal\ (left) and \hbe\ (right) lines of V4046~Sgr.
To emphasize variability, the average profile over the run is shown in red.
Orbital/rotation cycles (as listed in Table~1) are mentioned next to each profile.  }
\label{fig:bal}
\end{figure*}

Balmer lines are strongly in emission in V4046~Sgr, with widths large enough to blend the 
contribution of both system components into a single profile (at least for the first few lines 
of the series).  In the particular case of \hal\ and \hbe\ (see Fig.~\ref{fig:bal}), the 
total line equivalent widths of the time-averaged profiles are equal to 2,600~\kms\ (5.7~nm) 
and 475~\kms\ (0.77~nm), in line with previous reports \citep{Quast00, Stempels04}.  
The profiles are clearly variable with time but are dominated by intrinsic fluctuations 
and poorly repeat between consecutive rotational cycles (e.g., cycle 1.186 and 3.212).  
During our run, we observed in particular a strong 
short-lived, flare-like burst of emission (e.g., at cycle 2.386), with a conspicuous 
red-shifted emission component at velocities of up to 350~\kms\ suggesting that this event 
may reflect a short episode of enhanced mass accretion from the disc to the stars.  
This emission burst is also detected in the \hei\ $D_3$ line but not in the \caii\ IRT 
nor in the LSD profile of photospheric lines (see Fig.~\ref{fig:var} third panels).  
The width of Balmer lines ($\simeq$200~\kms\ for the full width at half maximum of \hal, 
\citealt{Quast00}, and $\simeq$400~\kms\ for the full width at 10\% height, see below) 
suggests that it traces free-falling material accelerated from large distances to the 
star, possibly from the inner rim of the circumbinary disc.  

Although small, we note potential signatures of rotational modulation in both \hal\ and 
\hbe, like for instance the recurrent pseudo-absorption episode detected in the red wing of 
both lines, at a mean velocity of about 200~\kms\ and close to phase 0.8 (at cycles 0.755 
and 2.799).  Similar (though usually deeper) absorption events are commonly observed in 
Balmer lines of cTTSs \citep[e.g.,][]{Bouvier07b, Donati08, Donati10b, Donati11} and are 
usually interpreted as evidence for accretion veils anchored within the surrounding disc and 
periodically intersecting the line of sight as the star rotates.  In the particular case of 
V4046~Sgr however, these episodes are only tracing pseudo-absorption (i.e., a lack of emission 
with respect to the mean profile) rather than true absorption (below the continuum level) and 
could also potentially be attributed to projection effects in a rotating non-axisymmetric 
magnetospheric accretion structure.  Data with denser phase coverage are necessary to 
address this issue in more details.  

From the average equivalent widths of the \caii\ IRT, \hei\ and \hbe\ lines, we 
derive logarithmic line fluxes (with respect to the luminosity of the Sun \lsun) for each star 
equal to $-5.0$, $-5.0$ and $-3.8$ respectively\footnote{To derive line fluxes from normalized 
equivalent widths, we approximate the continuum level by a Planck function at the temperature of 
the stellar photosphere.  Results are found to be compatible with those in the published literature 
\citep[e.g.,][]{Mohanty05} within better than 0.1~dex.  In the particular case of V4046~Sgr, we 
further assume that both stars exhibit more or less the same emission flux, i.e., that the total 
emission fluxes measured in the spectra reflect (within a factor of $\simeq$2) that of each 
system component (in rough agreement with what \caii\ IRT lines show, see Sec~\ref{sec:mod}).}, 
implying logarithmic accretion luminosities (with respect to \lsun) of $-2.2$, $-2.0$ and $-1.9$ 
respectively \citep[using empirical correlations from][]{Fang09}.  As estimates from the two main 
accretion proxies (\caii\ IRT and \hei\ $D_3$ lines) agree with each other and with the overall 
mean, we can safely conclude that the average logarithmic accretion luminosity of each component of 
V4046~Sgr is $-2.0\pm0.3$ and thus that the average logarithmic mass accretion rate of each star 
in V4046~Sgr (in \mspy) is equal to $-9.3\pm0.3$, with only limited fluctuations around this mean.  
Our estimate is in good agreement with that of \citet{Curran11}, equal to $-9.22\pm0.23$ and also 
derived from optical lines (though from fully independent data and a different analysis).  

Mass accretion rates can in principle also be estimated (though less accurately) through the full 
width of H$\alpha$ at 10\% height \citep[e.g.,][]{Natta04, Cieza10}.  In the case of V4046~Sgr, 
\hal\ shows a full width of 420~\kms\ in average, implying a logarithmic mass accretion rate            
estimate of $-8.8\pm0.6$ (in \mspy).  This is larger (though not very significantly) than the 
estimate derived from emission fluxes and in agreement with the results of \citet[][]{Curran11};  
at this stage, the origin of this potential discrepancy is unclear.

\section{Magnetic modelling}
\label{sec:mod}

\subsection{Overview of the method}

As for the previous papers, we aim at recovering the large-scale field topology of V4046~Sgr, 
as well as the distribution of surface cool spots and of chromospheric accretion regions, 
from the collected set of LSD and \caii\ IRT profiles presented and described in Secs.~\ref{sec:obs} 
and \ref{sec:var}.  We again apply our new modelling technique \citep[detailed extensively 
in][]{Donati10b, Donati11}, with the essential difference that we are now aiming at mapping the two 
stars of the binary system simultaneously.  

Following the principles of maximum entropy, our code automatically and simultaneously derives the 
simplest magnetic topologies, photospheric brightness images and accretion-powered \caii\ emission 
maps compatible with the series of rotationally modulated Stokes $I$ and $V$ LSD and \caii\ IRT 
profiles.  The reconstruction process is iterative and proceeds by comparing at each step the synthetic
Stokes $I$ and $V$ profiles corresponding to the current images with those of the observed data set.
The magnetic field of each star is described through its poloidal and toroidal components expressed
as spherical-harmonics (SH) expansions \citep{Donati06b}.  The spatial distributions of photospheric
brightness (with respect to the quiet photosphere) and those of accretion-powered \caii\ emission (in 
excess of and with respect to that produced by the quiet chromosphere) are modelled as series of 
independent pixels (typically a few thousand) on a grid covering the visible surfaces of the stars 
(with spots in the brightness images assumed to be darker/cooler than the quiet photospheres and 
spots in the accretion-powered \caii\ emission maps supposed to be brighter than the quiet 
chromospheres).

Synthetic profiles are computed by summing up the elementary spectral contributions from all image
pixels over the visible hemispheres of both stars, taking into account the relevant local parameters 
of the corresponding grid cells (e.g., brightness, accretion-powered excess emission, magnetic field 
strength and orientation, radial velocity, limb angle, projected area).  Since the problem is partly 
ill-posed, we stabilise the inversion process by using an entropy criterion (applied to the SH
coefficients and to the brightness/excess emission image pixels) aimed at selecting the field
topologies and images with minimum information among all those compatible with the data.
The relative weights attributed to the various SH modes can be imposed, e.g., for purposely
producing antisymmetric or symmetric field topologies with respect to the centre of the star
\citep[by favouring odd or even SH modes,][]{Donati07, Donati08}.
More details concerning the specific description of local profiles used in the model can be 
found in \citet{Donati10b}.  

\subsection{Modelling V4046~Sgr}

Given the relatively low level of intrinsic variability in the LSD photospheric and \caii\ 
IRT emission lines of V4046~Sgr (see Sec.~\ref{sec:var}), we applied our imaging model directly 
to the original profiles without passing them through our usual filtering procedure \citep[aimed 
at retaining rotational modulation only, see, e.g.,][]{Donati10b, Donati11}.  
We assume that $i=35$ (see Sec.~\ref{sec:v4046}), with values of $i$ ranging from 30\degr\ to 
40\degr\ yielding virtually identical results. 
We further assume that both stars rotate as solid bodies given the relatively limited coverage and 
sparse sampling of our data set;  assuming Sun-like differential rotation (both in sign and strength, 
i.e., inducing maximum phase delays of $\pm3$\% for the equator and pole with respect to the orbital cycle 
over the full time span of our observations) generates almost identical images, as expected from the 
limited spatial resolution available (see below).  

The parameters describing the local line profiles on both stars of V4046~Sgr are assumed to be 
very similar to those used in the previous studies \citep{Donati10b, Donati11}.  
In particular, the emission profile scaling factor $\epsilon$, describing the emission 
enhancement of accretion regions over the quiet chromosphere, is again set to $\epsilon=10$.  
The local filling factor $\psi$, describing the relative proportion of magnetic areas at any 
given point of the stellar surface and set here to $\psi=1$, has negligible impact on the result 
given the weak fields detected on V4046~Sgr.  

\begin{figure*}
\vspace{-2mm}
\hbox{\includegraphics[scale=0.7]{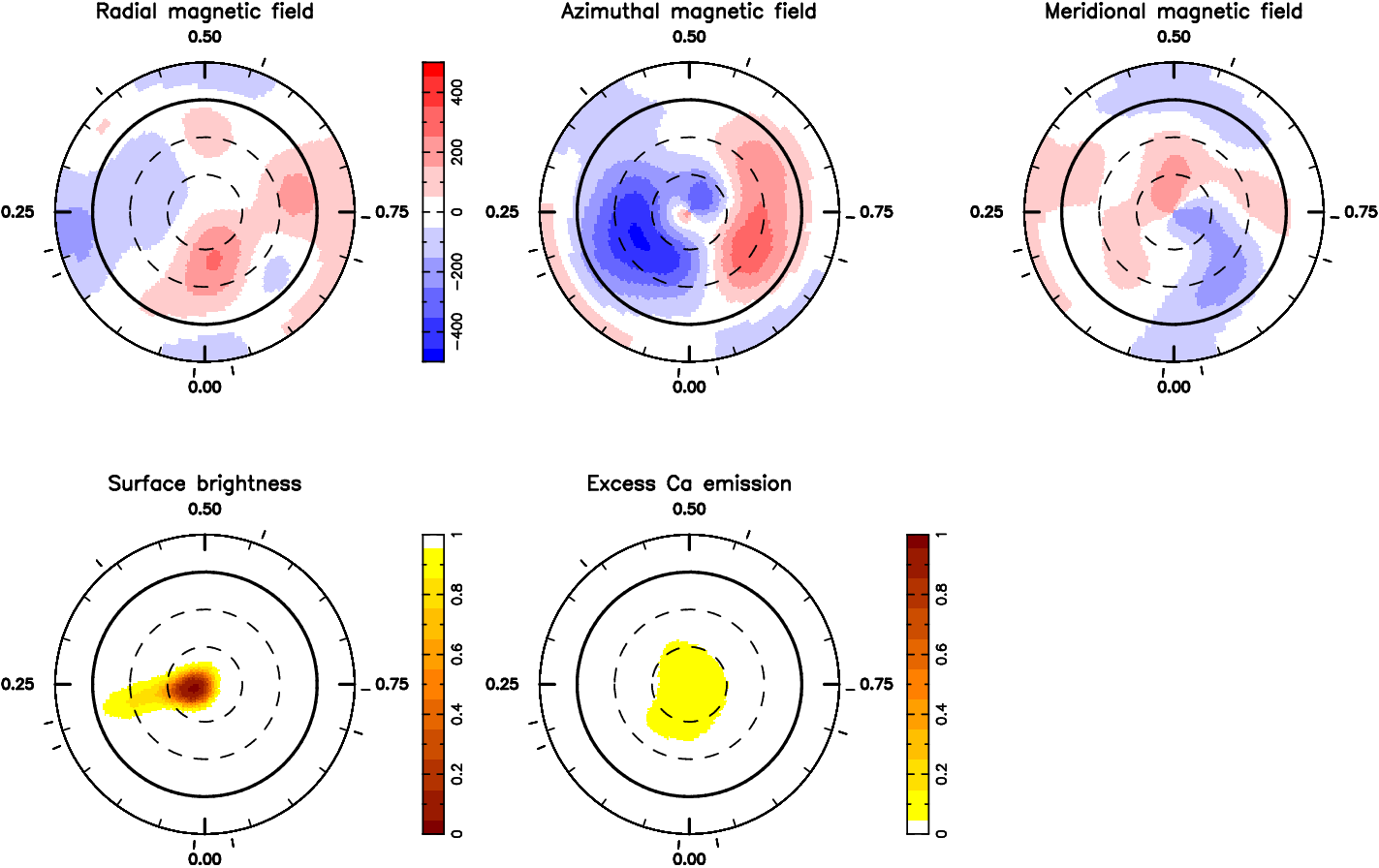}}
\vspace{8mm}
\hbox{\includegraphics[scale=0.7]{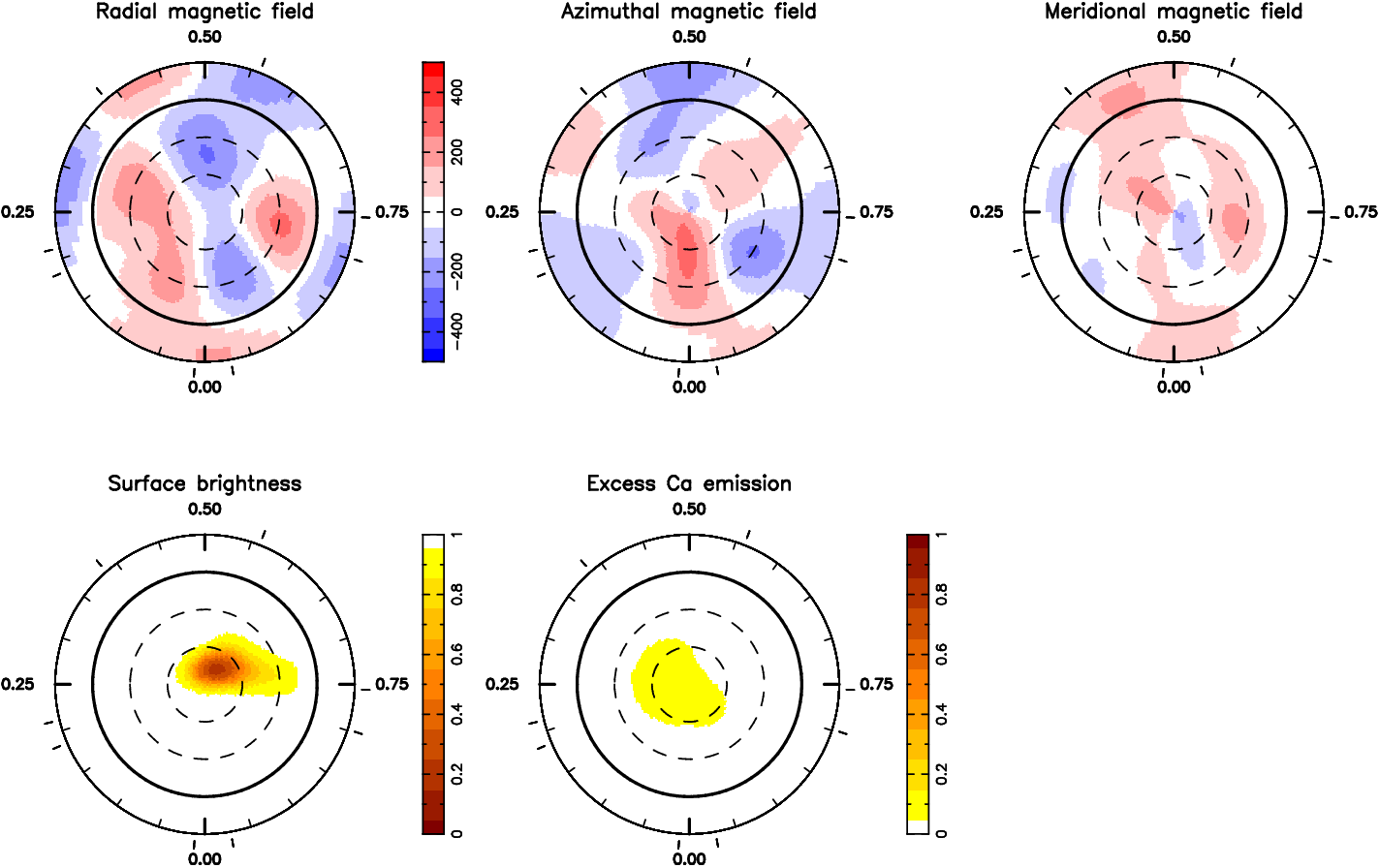}}
\caption[]{Maps of the radial, azimuthal and meridional components of the magnetic field $\bf B$
(first and third rows, left to right panels respectively), photospheric brightness and excess
\caii\ IRT emission (second and fourth rows, first and second panels respectively) at the
surface of the primary (top two rows) and secondary (bottom two rows) components of V4046~Sgr.  
Magnetic fluxes are labelled in G;  local photospheric brightness (normalized to that of the quiet
photosphere) varies from 1 (no spot) to 0 (no light);  local excess \caii\ emission varies from 0
(no excess emission) to 1 (excess emission covering 100\% of the local grid cell, assuming an
intrinsic excess emission of 10$\times$ the quiet chromospheric emission).
In all panels, the star is shown in flattened polar projection down to latitudes of $-30\degr$,
with the equator depicted as a bold circle and parallels as dashed circles.  Radial ticks around
each plot indicate phases of observations. The companion star faces phase 0.181 and 0.681 (i.e., 
the orbital phases of second and first conjunction) for the primary and secondary stars respectively. }
\label{fig:map}
\end{figure*}

\begin{figure*}
\center{
\includegraphics[scale=0.58,angle=-90]{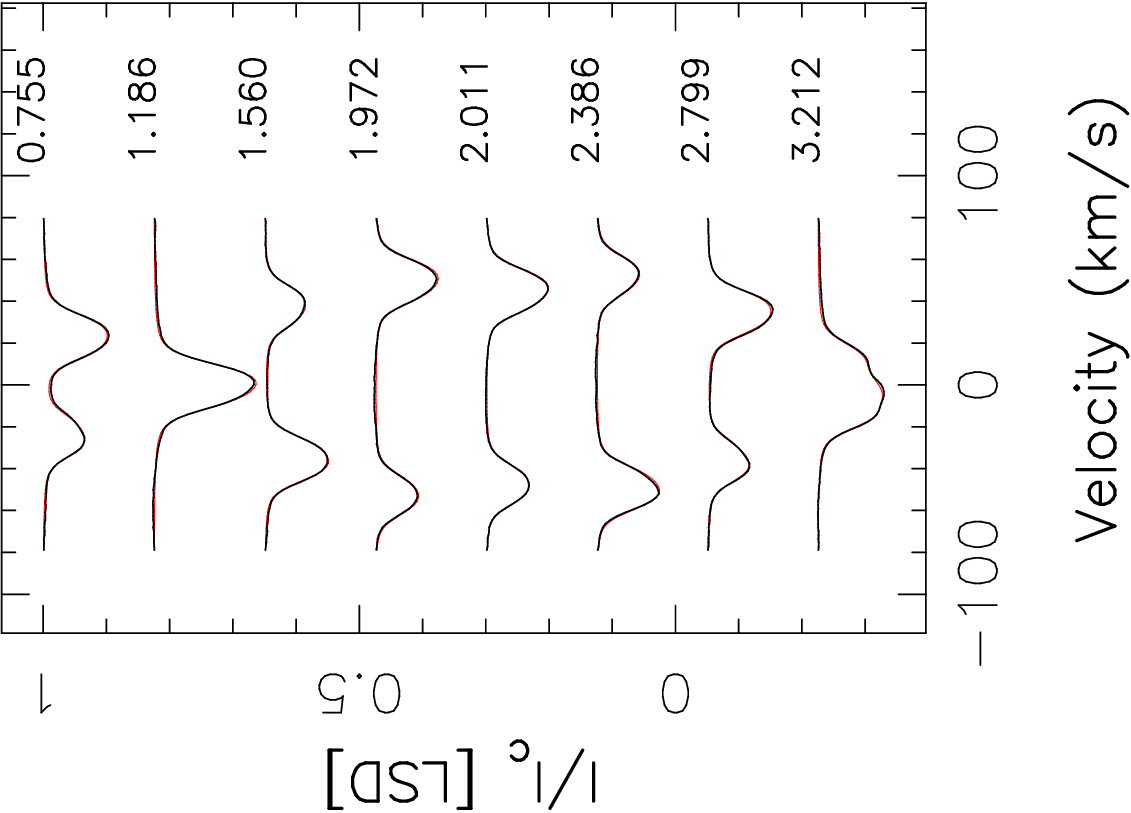}\hspace{4mm}
\includegraphics[scale=0.58,angle=-90]{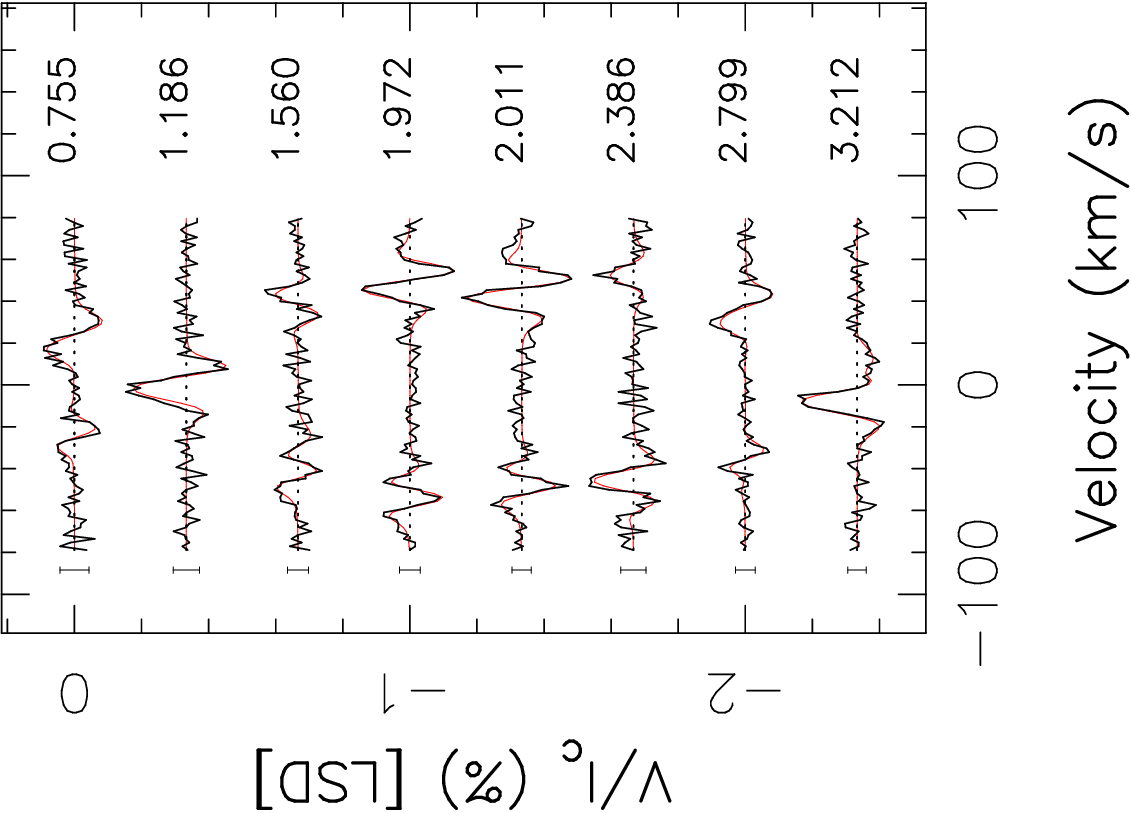}}
\vspace{6mm}
\center{
\includegraphics[scale=0.58,angle=-90]{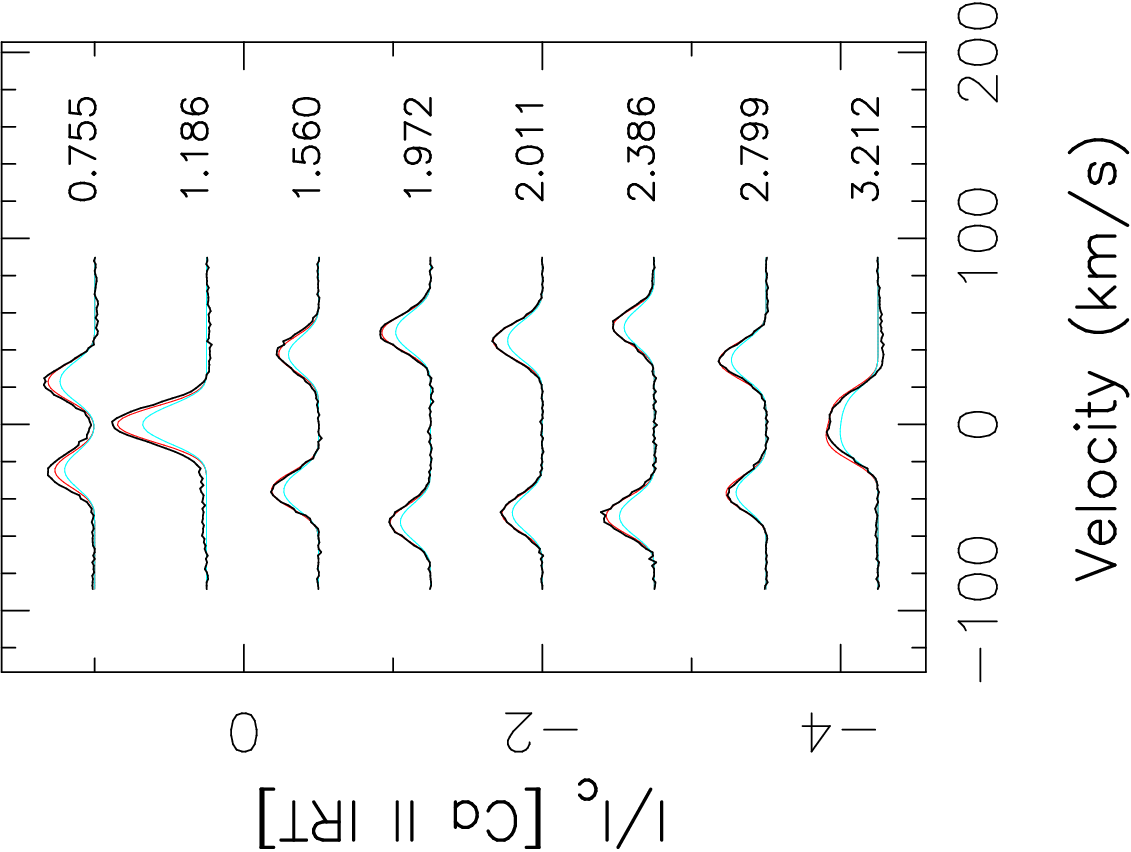}\hspace{4mm}
\includegraphics[scale=0.58,angle=-90]{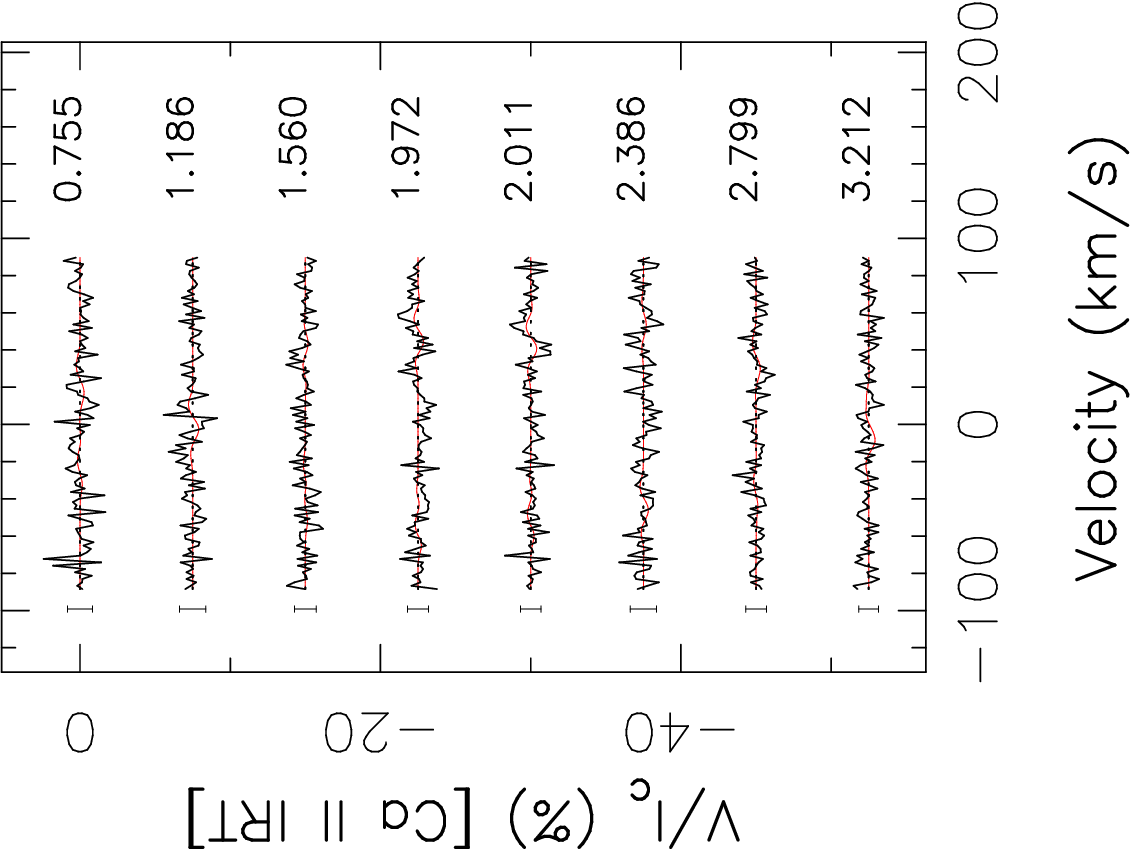}}
\caption[]{Maximum-entropy fit (thin red line) to the observed (thick black line) Stokes $I$ and
Stokes $V$ LSD photospheric profiles (top panels) and \caii\ IRT profiles (bottom panels)
of V4046~Sgr.  The light-blue curve in the bottom left panel shows the (constant)
contribution of the quiet chromosphere to the Stokes $I$ \caii\ profiles.
Orbital/rotational cycles and 3$\sigma$ error bars (for Stokes $V$ profiles) are also shown next to each
profile.  First and second orbital conjunctions occur at phases 0.681 and 0.181 respectively.  }
\label{fig:fit}
\end{figure*}

The magnetic, brightness and accretion maps we reconstruct for both stars of V4046~Sgr are 
shown in Fig.~\ref{fig:map}, with corresponding fits to the data shown in Fig.~\ref{fig:fit}.  
The SH expansion describing the field was limited to terms with $\ell\leq5$; first attempts 
with $\ell=8$ indicate that little power is reconstructed in higher order modes ($\ell\ge6$), 
reflecting essentially the limited spatial information accessible to Doppler tomography at 
moderate \vsini\ values.  
For this modelling, we assumed (as in previous studies) that the fields of both stars are 
antisymmetric (with respect to the stellar centres);  this assumption however has little impact 
on the reconstructed magnetic maps over the visible regions of the stellar surfaces.  
Our fits reproduce the data down to the noise level, i.e., to a unit reduced chi-square 
\chisqr\ starting from an initial \chisqr\ of 3.2.  

As a by-product, we obtain accurate estimates of the system orbital elements, and in particular 
of the phase of first conjunction $\phi_0=0.6809\pm0.0004$, of the velocity amplitude of the 
primary and secondary components $K_1=51.4\pm0.2$ and $K_2=54.3\pm0.2$~\kms, and of the system 
radial velocity $\gamma=-5.7\pm0.2$~\kms.  These estimates are in agreement with those of
\citet{Quast00} except for the latter ($\gamma$) whose significant variation suggests the 
presence of a third body in the system.  Note in particular that our estimates of $K_1$ and 
$K_2$ are mostly free of contamination from cool surface spots \citep[as opposed to those 
of][]{Quast00} since they are derived as part of the imaging process;  we suggest that the 
small difference between our estimates and those of \citet{Quast00} (about 0.5~\kms) 
mostly reflects the RV signatures of cool surface spots (see Sec.~\ref{sec:var} and 
Fig.~\ref{fig:var} left panel second row).    

We also find that \caii\ lines are redshifted by 
about 0.7~\kms\ with respect to rest frames of both stars, a typical value for cTTSs.  
In addition, we derive estimates of the projected equatorial rotation velocities \vsini\ for 
both stars, equal to $13.5\pm0.5$ and $12.5\pm0.5$~\kms\ for the primary and secondary component 
respectively, as well as a measurement of the magnitude contrast between both system 
components at the average wavelengths of the LSD and \caii\ IRT profiles (620 and 850~nm), 
equal to $0.52\pm0.10$ and $0.22\pm0.15$ respectively\footnote{We speculate that the smaller 
contrast observed in the \caii\ IRT lines reflects the fact that the cooler secondary star is 
comparatively more active and/or generating more accretion-powered emission than the primary 
star, by typically 35\%. }.  

\subsection{Modelling results}

The reconstructed large-scale magnetic topologies of both cTTSs in the V4046~Sgr system 
are weak and multipolar, as expected from the low amplitude and complex shape of the
detected Zeeman signatures.  
On the primary star, the magnetic field has an average surface intensity of $\simeq$230~G, 
with the azimuthal field component dominating the others and reaching maximum strength 
($\simeq$500~G) in conspicuous arc-like structures around the pole;  as a consequence, 
the toroidal field component is found to be very significant, totalling as much magnetic 
energy as the poloidal component.  The poloidal field component is mostly non-axisymmetric, 
with less than 10\% of the energy concentrating in modes with $m<\ell/2$;  the dipole 
(i.e., $\ell=1$) component has a polar strength of $\simeq$100~G and is highly tilted 
(at about 60\degr) with respect to the rotation axis (pointing towards phase 0.80).  
On the secondary star, the magnetic field has a comparable average surface intensity 
(of $\simeq$170~G), with the radial and azimuthal field components of roughly comparable 
strengths and topologies.  The toroidal field component is weaker than in the 
primary star though still significant, totalling only about 15\% of the reconstructed magnetic energy;  the 
poloidal field component is once more mostly non-axisymmetric, with again less than 
10\% of the energy concentrating in modes with $m<\ell/2$;  the dipole component is even
weaker ($\simeq$70~G) than on the primary star and is roughly perpendicular to the 
rotation axis (pointing towards phase 0.1) and to the dipole moment of the primary 
star.  

The photospheric brightness maps we derive for both stars of V4046~Sgr feature one dark 
spot close to the pole and each covering $\simeq$2\% of the stellar surface.  Both spots 
are slightly off-centered (by about $\simeq$10\degr) towards phase $\simeq$0.2 for the 
primary star and phase $\simeq$0.7 for the secondary star, i.e., towards the hemisphere 
that faces the companion star (best viewed from Earth at the orbital phases of the 
second and first conjunctions, respectively equal to 0.181 and 0.681, see Sec.~\ref{sec:obs}).  
This is fully consistent with the observed modulation of the LSD profiles of 
photospheric lines as reported in Sec.~\ref{sec:var}, showing that the primary is 
brighter/fainter than the secondary around phases 0.7/0.2 respectively.  No obvious correlation between 
the magnetic and brightness maps, apart from the fact that the (incomplete) counterclockwise 
ring of azimuthal/toroidal field detected at high latitudes on the primary star is more 
or less surrounding the corresponding cool polar spot (see Fig.~\ref{fig:map} upper panels).  

The reconstructed maps of accretion-powered excess emission of both stars mainly show 
an extended, low-contrast region of enhanced emission centred onto the pole, each covering 
$\simeq$1\% of the stellar surface.  Again, this is in agreement with the moderate 
strength and small (largely insignificant) modulation of the \caii\ emission profiles of 
V4046~Sgr as reported in Sec.~\ref{sec:var}.  As for the brightness maps, we see no obvious 
connection between these features and the corresponding magnetic maps.  
We suspect that these extended, low-contrast regions of excess emission centred on the pole 
may be tracing even more extended features potentially covering most of the star, but with 
low latitudes truncated by tomographic imaging - the maximum entropy criterion naturally 
filtering out areas of lowest visibility (i.e., lowest latitudes) when phase sampling and 
spatial resolution are both moderate (like in the present case).  
Simulations involving fake stars with properties similar to those of V4046~Sgr (regarding 
\vsini\ and $i$ in particular) and with low-contrast excess emission evenly distributed 
over the whole surfaces confirm that the reconstructed emission features (centred on the pole) 
extend no further than latitude 45\degr\ typically (i.e., only slightly larger than what we 
recover in the case of V4046~Sgr)  when the quality and sampling of the spectra are similar to 
those of our observations.  
Our maps therefore suggest that excess emission likely extends over a large area on both 
stars of V4046~Sgr, with no more than a small residual excess in the polar regions;  
new observations with improved phase sampling are needed to validate more firmly this 
conclusion.

\section{Summary \& discussion}
\label{sec:dis}

Our paper presents new results regarding magnetospheric accretion processes in cTTSs;  
here we concentrate on the bright cTTS close binary V4046~Sgr, comprised of two 
$\simeq$12~Myr-old $\simeq$0.9~\msun\ stars separated by 0.041~AU in a 2.42~d-period 
circular orbit, and observed within the framework of the international multi-wavelength 
multi-site monitoring campaign organised for this object in 2009 September (involving 
in particular X-ray observations with XMM-Newton;  Argiroffi et al 2011, submitted).  
Spectropolarimetric data were collected with ESPaDOnS@CFHT over $\simeq$2.5 orbital/rotation 
cycles.  From these time-resolved spectropolarimetric data and using the 
latest version of our magnetic tomographic imaging code (extended to the case of close 
double-line binaries), we reconstruct maps of the large-scale magnetic field, of the
photospheric brightness and of the accretion-powered \caii\ IRT excess emission at
the surface of both the primary and secondary stars of V4046~Sgr.  

The large-scale fields of both stars are found to be weak and complex (with respect 
to those of other cTTSs of similar masses in particular), with average field strengths 
of typically $\simeq$200~G, a highly non-axisymmetric poloidal component, and a 
significant toroidal component;  in particular, the large-scale dipole field components 
are weak (100 and 70~G for the primary and secondary star respectively), highly tilted 
with respect to the rotation axis and perpendicular to each other.  This is 
radically different from the strong, mainly poloidal, mostly axisymmetric magnetic 
topologies of prototypical cTTSs such as BP~Tau or AA~Tau \citep{Donati08, Donati10b}, 
and similar to field configurations of more massive cTTSs \citep[e.g.,][]{Hussain09}. 
The field configurations of both cTTSs of V4046~Sgr are also quite similar to those 
found in HD~155555 (another slightly more massive close binary of the $\beta$~Pic moving 
group, see Sec.~\ref{sec:v4046}), also showing a weak and complex field with a highly 
non-axisymmetric poloidal component and a significant toroidal component 
\citep{Dunstone08a}.  

By analogy with dynamo-generated magnetic topologies of main-sequence stars, we speculate 
that this contrast reflects a drastic difference in the internal structure of the star, 
with fully-convective stars (those not too far below the full-convection limit at least) 
showing strong, mostly axisymmetric and poloidal magnetic configurations and 
partly-convective stars exhibiting much weaker and more complex poloidal fields and 
a significant toroidal component \citep{Donati09}.  Given their likely age of $\simeq$12~Myr, 
the components of V4046~Sgr should belong to the second category while BP~Tau and AA~Tau 
belong to the first, further strengthening the conclusion that magnetic fields of cTTSs 
have a dynamo origin \citep[e.g.,][]{Donati11}.  

Our images also show that both stars of V4046~Sgr host cool spots close to the pole, 
similar to (though smaller in size than) those mapped at the surface of HD~155555 
\citep{Dunstone08a};  we also find that, for each star in the system, these spots are 
slightly off-centred towards the hemisphere that faces the companion star, in 
agreement with the primary star being brighter and fainter than the secondary star 
(as judged from the primary to secondary line depth ratio, see Sec.~\ref{sec:var} and 
Fig.~\ref{fig:var}) at orbital phases of first and second conjunction respectively (i.e., 
phases 0.681 and 0.181 respectively).  
We find no clear correlation between the reconstructed brightness maps and corresponding magnetic 
topologies, despite both being reconstructed simultaneously in a fully self-consistent 
way;  in particular, the cool polar spots are apparently not coincident with 
one pole of a mainly axisymmetric magnetic topology as in most other cTTSs 
\citep{Donati08, Donati10b, Donati11}.  The only connection we can report between magnetic 
topologies and photospheric brightness maps is that the (incomplete) unipolar ring of 
counterclockwise azimuthal field detected on the primary star is encircling the cool 
polar spot;  this is very similar to what is reported for rapidly-rotating 
partly-convective solar-type stars \citep[e.g.,][]{Donati03a, Dunstone08a}.  

The maps of accretion-powered excess emission we recover for both stars from sets 
of \caii\ IRT profiles show a low-contrast, extended feature centred on the pole, 
directly reflecting the low emission and small modulation that the \caii\ IRT lines 
exhibit (see Sec.~\ref{sec:var} and Fig.~\ref{fig:var}).  In particular, these maps are 
less contrasted that those of most other cTTSs imaged to date \citep{Donati10b, Donati11} 
despite a similar mass accretion rate (of $-9.3\pm0.3$ in logarithmic scale 
and in \mspy, see Sec.~\ref{sec:var});  they also show no obvious resemblance 
with the corresponding magnetic maps.  Tomographic imaging simulations suggest 
that the parent distributions of excess emission at the surfaces of both stars of 
V4046~Sgr may potentially cover the whole surfaces (with signal from low latitudes 
being filtered out by the maximum entropy criterion) and no more than a low-contrast 
residual emission excess close to the pole.  

We suggest that this reflects that accretion onto each star of 
V4046~Sgr does not concentrate into a single specific region of the
visible hemisphere \citep[as it does on other cTTSs, e.g., ][]{Donati08, Donati10b, 
Donati11} but rather occurs simultaneously at several places over the whole stellar 
surface;  in each of the individual accretion sites, mass accretion is presumably 
not high enough to produce distinct features in the maps of accretion-powered 
emission excesses, especially given the limited spatial resolution achievable at 
the surfaces of both stars of V4046~Sgr.  Any residual low-contrast excess emission 
close to the pole may indicate that chromospheric emission is not homogeneous 
on both stars of V4046~Sgr (as originally assumed in the model) but is slightly 
stronger at high latitudes, where cool spots are observed to cluster at 
photospheric level.  

Distributed accretion is consistent with the fact that the dipolar components of 
the large-scale fields of both stars are weak and that the field topology within the 
magnetospheric gap (extending no further than 0.5--1~\rstar\ above the surface of
both components of V4046~Sgr) is complex, dominated by higher-order terms of 
the magnetic spherical expansions and thus featuring multiple poles over the visible 
hemisphere.  
The red-shifted pseudo-absorption episodes potentially observed in Balmer lines at 
specific phases (rather than throughout the whole rotation cycle, see Sec.~\ref{sec:var}) 
may provide more information about the spatial intermittency of accretion funnels, if 
further data (with much denser phase coverage) demonstrate that these episodes are 
reliable probes of accretion funnels crossing the line of sight in V4046~Sgr as well.  

Recent 2D and 3D numerical hydrodynamical simulations of mass accretion onto close cTTS 
binaries \citep{kaygorodov10, fateeva11, devalborro11} suggest that accretion may not 
directly proceed from the circumbinary disc to the stars;  more specifically, they find 
that local small-size discs may develop around both stars, with material flowing from the 
circumbinary disc to the small discs through the circumbinary gap (and in particular 
through the co-rotating outer Lagrange points), and forming a double-armed spiral 
structure (with one arm joining each local disc to the wider circumbinary disc) as well 
as a bridge connecting both local discs.  The local discs have predicted sizes similar 
to those of the corresponding Roche lobes, with radii of $\simeq$3~\rstar\ in the 
particular case of V4046~Sgr.  Observations of optically-thin Balmer lines however 
suggest that only the accretion streams actually exist in V4046~Sgr and not the local 
discs, as a possible result of them being magnetically disrupted by the large-scale 
fields of both stars \citep{devalborro11}.  

Self-consistent simulations including the effects of magnetic fields are not available yet.  
As of now, our results suggest that the large-scale fields of both stars, and in particular 
their dipolar components, are likely not strong enough to disrupt the predicted local discs 
around both system stars beyond a typical distance of 1~\rstar\ above the stellar surfaces.  
However, our data can neither directly confirm nor invalidate the existence of local accretion 
discs.  For instance, the \caii\ IRT emission components, although a 
useful tracer of the hot footpoints at the base of accretion funnels, show no evidence of 
accretion spots located beyond the surfaces of the system stars, e.g., where the spiral 
accretion streams supposedly meet the putative local accretion discs;  it is however not 
clear from the existing simulations whether such spots truly exist and, if they do, 
whether they are bright enough to be detected.  

The time-resolved contemporaneous (though not exactly simultaneous) X-ray observations 
of V4046~Sgr collected with XMM-Newton in the framework of our international campaign 
can provide useful constraints onto magnetospheres and accretion processes in a close 
cTTS binary like V4046~Sgr.  
They confirm in particular that the 
X-ray emission measure distribution of V4046~Sgr is similar to that of the evolved 
cTTS TW~Hya, featuring in particular a hard X-ray component of coronal origin 
(with plasma temperatures of up to 10~MK) and a soft X-ray one (with an average temperature of 
about 3~MK) presumably probing the accretion shocks at the base of accretion funnels 
(Argiroffi et al.\ 2011, submitted).  That TW~Hya hosts a strong and mostly axisymmetric 
large-scale magnetic field \citep{Donati11b} drastically different 
from those of both stars of V4046~Sgr makes this similarity very intriguing and even more 
interesting to study.  

At least three major flares in the hot X-ray component were monitored (including both the 
rise and decay phases) during these observations, suggesting 
flaring loops with maximum sizes comparable to the stellar radii (Maggio et al.\ 2011, 
in preparation);  this is in good agreement with our independent conclusion that the 
magnetospheres of both stars in V4046~Sgr do not extend beyond 1~\rstar\ above the 
surface of the star.  Since rotation and orbital motion are locked in V4046~Sgr, and 
given that the distance between both system components ($\simeq$8.8~\rstar) is much 
larger than the reported loop sizes and magnetospheres, this flaring activity is unlikely 
caused by interacting magnetospheres \citep[as proposed for DQ~Tau,][]{Getman11} 
but is rather attributable to conventional coronal activity similar to that observed on 
most active Sun-like stars and presumably triggered by differential rotation at the surface 
of both stars mostly (rather than to rapid changes in the surface magnetic topologies 
themselves).  


Obviously, more work is required, both on the observational and theoretical sides, to 
understand how magnetospheric accretion processes work in the case of close cTTSs 
binaries, and in particular how they differ from those occurring in single cTTSs.  
Coordinated multi-wavelength campaigns like that organized for V4046~Sgr and described 
in the present and forthcoming papers, are crucial in this respect, and need to be 
carried out for binary stars of different masses, rotation/orbital rates and ages.

\section*{Acknowledgements}
We thank an anonymous referee for comments that improved the manuscript.  
This paper is based on 
observations obtained at the Canada-France-Hawaii Telescope (CFHT), operated by the National 
Research Council of Canada, the Institut National des Sciences de l'Univers of the Centre 
National de la Recherche Scientifique of France and the University of Hawaii. 
The ``Magnetic Protostars and Planets'' (MaPP) project is supported by the 
funding agencies of CFHT and TBL (through the allocation of telescope time) 
and by CNRS/INSU in particular, as well as by the French ``Agence Nationale 
pour la Recherche'' (ANR).  SHPA acknowledges financial
support from Fapemig, CAPES and COFECUB.

\bibliography{v4046}

\begin{thebibliography}{}

\bibitem[\protect\citeauthoryear{{Andr{\'e}}, {Basu} \& {Inutsuka}}{{Andr{\'e}}
  et~al.}{2009}]{Andre09}
{Andr{\'e}} P.,  {Basu} S.,    {Inutsuka} S.,  2009, {The formation and
  evolution of prestellar cores}.
Cambridge University Press, p.~254

\bibitem[\protect\citeauthoryear{{Applegate}}{{Applegate}}{1992}]{Applegate92}
{Applegate} J.~H.,  1992, \apj, 385, 621

\bibitem[\protect\citeauthoryear{{Bouvier}, {Alencar}, {Boutelier}, {Dougados},
  {Balog}, {Grankin}, {Hodgkin}, {Ibrahimov}, {Kun}, {Magakian} \&
  {Pinte}}{{Bouvier} et~al.}{2007b}]{Bouvier07b}
{Bouvier} J.,  {Alencar} S.~H.~P.,  {Boutelier} T.,  {Dougados} C.,  {Balog}
  Z.,  {Grankin} K.,  {Hodgkin} S.~T.,  {Ibrahimov} M.~A.,  {Kun} M.,
  {Magakian} T.~Y.,    {Pinte} C.,  2007b, \aap, 463, 1017

\bibitem[\protect\citeauthoryear{{Bouvier}, {Alencar}, {Harries}, {Johns-Krull}
  \& {Romanova}}{{Bouvier} et~al.}{2007a}]{Bouvier07}
{Bouvier} J.,  {Alencar} S.~H.~P.,  {Harries} T.~J.,  {Johns-Krull} C.~M.,
  {Romanova} M.~M.,  2007a, in {Reipurth} B.,  {Jewitt} D.,   {Keil} K.,  eds,
  Protostars and Planets V {Magnetospheric Accretion in Classical T Tauri
  Stars}.
pp 479--494

\bibitem[\protect\citeauthoryear{{Cieza}, {Schreiber}, {Romero}, {Mora},
  {Merin}, {Swift}, {Orellana}, {Williams}, {Harvey} \& {Evans}}{{Cieza}
  et~al.}{2010}]{Cieza10}
{Cieza} L.~A.,  {Schreiber} M.~R.,  {Romero} G.~A.,  {Mora} M.~D.,  {Merin} B.,
   {Swift} J.~J.,  {Orellana} M.,  {Williams} J.~P.,  {Harvey} P.~M.,
  {Evans} N.~J.,  2010, \apj, 712, 925

\bibitem[\protect\citeauthoryear{{Curran}, {Argiroffi}, {Sacco}, {Orlando},
  {Peres}, {Reale} \& {Maggio}}{{Curran} et~al.}{2011}]{Curran11}
{Curran} R.~L.,  {Argiroffi} C.,  {Sacco} G.~G.,  {Orlando} S.,  {Peres} G.,
  {Reale} F.,    {Maggio} A.,  2011, \aap, 526, A104

\bibitem[\protect\citeauthoryear{{da Silva}, {Torres}, {de La Reza}, {Quast},
  {Melo} \& {Sterzik}}{{da Silva} et~al.}{2009}]{daSilva09}
{da Silva} L.,  {Torres} C.~A.~O.,  {de La Reza} R.,  {Quast} G.~R.,  {Melo}
  C.~H.~F.,    {Sterzik} M.~F.,  2009, \aap, 508, 833

\bibitem[\protect\citeauthoryear{{de Val-Borro}, {Gahm}, {Stempels} \&
  {Pepli{\'n}ski}}{{de Val-Borro} et~al.}{2011}]{devalborro11}
{de Val-Borro} M.,  {Gahm} G.~F.,  {Stempels} H.~C.,    {Pepli{\'n}ski} A.,
  2011, \mnras, 413, 2679

\bibitem[\protect\citeauthoryear{{Donati}, {Bouvier}, {Walter}, {Gregory},
  {Skelly}, {Hussain}, {Flaccomio}, {Argiroffi}, {Grankin}, {Jardine},
  {M{\'e}nard}, {Dougados} \& {Romanova}}{{Donati} et~al.}{2011a}]{Donati11}
{Donati} J.,  {Bouvier} J.,  {Walter} F.~M.,  {Gregory} S.~G.,  {Skelly} M.~B.,
   {Hussain} G.~A.~J.,  {Flaccomio} E.,  {Argiroffi} C.,  {Grankin} K.~N.,
  {Jardine} M.~M.,  {M{\'e}nard} F.,  {Dougados} C.,    {Romanova} M.~M.,
  2011a, \mnras, 412, 2454

\bibitem[\protect\citeauthoryear{{Donati}, {Gregory}, {Alencar}, {Bouvier},
  {Hussain}, {Skelly}, {Dougados}, {Jardine}, {Menard}, {Romanova}, {Unruh} \&
  {the MaPP collaboration}}{{Donati} et~al.}{2011b}]{Donati11b}
{Donati} J.,  {Gregory} S.,  {Alencar} S.,  {Bouvier} J.,  {Hussain} G.,
  {Skelly} M.,  {Dougados} C.,  {Jardine} M.,  {Menard} F.,  {Romanova} M.,
  {Unruh} Y.,    {the MaPP collaboration} 2011, MNRAS in press (arXiv:1106.4162) 

\bibitem[\protect\citeauthoryear{{Donati} \& {Landstreet}}{{Donati} \&
  {Landstreet}}{2009}]{Donati09}
{Donati} J.,  {Landstreet} J.~D.,  2009, \araa, 47, 333

\bibitem[\protect\citeauthoryear{{Donati}, {Skelly}, {Bouvier}, {Gregory},
  {Grankin}, {Jardine}, {Hussain}, {M{\'e}nard}, {Dougados}, {Unruh},
  {Mohanty}, {Auri{\`e}re}, {Morin} \& {Far{\`e}s}}{{Donati}
  et~al.}{2010}]{Donati10b}
{Donati} J.,  {Skelly} M.~B.,  {Bouvier} J.,  {Gregory} S.~G.,  {Grankin}
  K.~N.,  {Jardine} M.~M.,  {Hussain} G.~A.~J.,  {M{\'e}nard} F.,  {Dougados}
  C.,  {Unruh} Y.,  {Mohanty} S.,  {Auri{\`e}re} M.,  {Morin} J.,
  {Far{\`e}s} R.,  2010, \mnras, 409, 1347

\bibitem[\protect\citeauthoryear{{Donati}}{{Donati}}{2003}]{Donati03}
{Donati} J.-F.,  2003, in {Trujillo-Bueno} J.,  {Sanchez Almeida} J.,  eds,
  Astronomical Society of the Pacific Conference Series Vol.~307 of
  Astronomical Society of the Pacific Conference Series, {ESPaDOnS: An Echelle
  SpectroPolarimetric Device for the Observation of Stars at CFHT}.
p.~41

\bibitem[\protect\citeauthoryear{{Donati}, {Cameron}, {Semel}, {Hussain},
  {Petit}, {Carter}, {Marsden}, {Mengel}, {Lopez Ariste}, {Jeffers} \&
  {Rees}}{{Donati} et~al.}{2003}]{Donati03a}
{Donati} J.-F.,  {Cameron} A.,  {Semel} M.,  {Hussain} G.,  {Petit} P.,
  {Carter} B.,  {Marsden} S.,  {Mengel} M.,  {Lopez Ariste} A.,  {Jeffers} S.,
    {Rees} D.,  2003, \mnras, 345, 1145

\bibitem[\protect\citeauthoryear{{Donati}, {Howarth}, {Jardine}, {Petit},
  {Catala}, {Landstreet}, {Bouret}, {Alecian}, {Barnes}, {Forveille}, {Paletou}
  \& {Manset}}{{Donati} et~al.}{2006}]{Donati06b}
{Donati} J.-F.,  {Howarth} I.~D.,  {Jardine} M.~M.,  {Petit} P.,  {Catala} C.,
  {Landstreet} J.~D.,  {Bouret} J.-C.,  {Alecian} E.,  {Barnes} J.~R.,
  {Forveille} T.,  {Paletou} F.,    {Manset} N.,  2006, \mnras, 370, 629

\bibitem[\protect\citeauthoryear{{Donati}, {Jardine}, {Gregory}, {Petit},
  {Bouvier}, {Dougados}, {M{\'e}nard}, {Cameron}, {Harries}, {Jeffers} \&
  {Paletou}}{{Donati} et~al.}{2007}]{Donati07}
{Donati} J.-F.,  {Jardine} M.~M.,  {Gregory} S.~G.,  {Petit} P.,  {Bouvier} J.,
   {Dougados} C.,  {M{\'e}nard} F.,  {Cameron} A.~C.,  {Harries} T.~J.,
  {Jeffers} S.~V.,    {Paletou} F.,  2007, \mnras, 380, 1297

\bibitem[\protect\citeauthoryear{{Donati}, {Jardine}, {Gregory}, {Petit},
  {Paletou}, {Bouvier}, {Dougados}, {M{\'e}nard}, {Cameron}, {Harries},
  {Hussain}, {Unruh}, {Morin}, {Marsden}, {Manset}, {Auri{\`e}re}, {Catala} \&
  {Alecian}}{{Donati} et~al.}{2008b}]{Donati08}
{Donati} J.-F.,  {Jardine} M.~M.,  {Gregory} S.~G.,  {Petit} P.,  {Paletou} F.,
   {Bouvier} J.,  {Dougados} C.,  {M{\'e}nard} F.,  {Cameron} A.~C.,  {Harries}
  T.~J.,  {Hussain} G.~A.~J.,  {Unruh} Y.,  {Morin} J.,  {Marsden} S.~C.,
  {Manset} N.,  {Auri{\`e}re} M.,  {Catala} C.,    {Alecian} E.,  2008b, \mnras,
  386, 1234

\bibitem[\protect\citeauthoryear{{Donati}, {Moutou}, {Far{\`e}s}, {Bohlender},
  {Catala}, {Deleuil}, {Shkolnik}, {Cameron}, {Jardine} \& {Walker}}{{Donati}
  et~al.}{2008a}]{Donati08b}
{Donati} J.-F.,  {Moutou} C.,  {Far{\`e}s} R.,  {Bohlender} D.,  {Catala} C.,
  {Deleuil} M.,  {Shkolnik} E.,  {Cameron} A.~C.,  {Jardine} M.~M.,    {Walker}
  G.~A.~H.,  2008a, \mnras, 385, 1179

\bibitem[\protect\citeauthoryear{{Donati}, {Semel}, {Carter}, {Rees} \&
  {Collier Cameron}}{{Donati} et~al.}{1997}]{Donati97b}
{Donati} J.-F.,  {Semel} M.,  {Carter} B.~D.,  {Rees} D.~E.,    {Collier
  Cameron} A.,  1997, \mnras, 291, 658

\bibitem[\protect\citeauthoryear{{Dunstone}, {Hussain}, {Collier Cameron},
  {Marsden}, {Jardine}, {Stempels}, {Ramirez Velez} \& {Donati}}{{Dunstone}
  et~al.}{2008}]{Dunstone08a}
{Dunstone} N.~J.,  {Hussain} G.~A.~J.,  {Collier Cameron} A.,  {Marsden} S.~C.,
   {Jardine} M.,  {Stempels} H.~C.,  {Ramirez Velez} J.~C.,    {Donati} J.-F.,
  2008, \mnras, 387, 481

\bibitem[\protect\citeauthoryear{{Fang}, {van Boekel}, {Wang}, {Carmona},
  {Sicilia-Aguilar} \& {Henning}}{{Fang} et~al.}{2009}]{Fang09}
{Fang} M.,  {van Boekel} R.,  {Wang} W.,  {Carmona} A.,  {Sicilia-Aguilar} A.,
    {Henning} T.,  2009, \aap, 504, 461

\bibitem[\protect\citeauthoryear{{Fateeva}, {Bisikalo}, {Kaygorodov} \&
  {Sytov}}{{Fateeva} et~al.}{2011}]{fateeva11}
{Fateeva} A.~M.,  {Bisikalo} D.~V.,  {Kaygorodov} P.~V.,    {Sytov} A.~Y.,
  2011, \apss, p.~29

\bibitem[\protect\citeauthoryear{{Favata} \& {Micela}}{{Favata} \&
  {Micela}}{2003}]{Favata03}
{Favata} F.,  {Micela} G.,  2003, \ssr, 108, 577

\bibitem[\protect\citeauthoryear{{Feigelson} \& {Montmerle}}{{Feigelson} \&
  {Montmerle}}{1999}]{Feigelson99}
{Feigelson} E.~D.,  {Montmerle} T.,  1999, \araa, 37, 363

\bibitem[\protect\citeauthoryear{{Getman}, {Broos}, {Salter}, {Garmire} \&
  {Hogerheijde}}{{Getman} et~al.}{2011}]{Getman11}
{Getman} K.~V.,  {Broos} P.~S.,  {Salter} D.~M.,  {Garmire} G.~P.,
  {Hogerheijde} M.~R.,  2011, \apj, 730, 6

\bibitem[\protect\citeauthoryear{{Gregory}, {Jardine}, {Gray} \&
  {Donati}}{{Gregory} et~al.}{2010}]{Gregory10}
{Gregory} S.~G.,  {Jardine} M.,  {Gray} C.~G.,    {Donati} J.,  2010, Reports
  on Progress in Physics, 73, 126901

\bibitem[\protect\citeauthoryear{{G{\"u}del} \& {Naz{\'e}}}{{G{\"u}del} \&
  {Naz{\'e}}}{2009}]{Gudel09}
{G{\"u}del} M.,  {Naz{\'e}} Y.,  2009, \aapr, 17, 309

\bibitem[\protect\citeauthoryear{{G{\"u}nther}, {Liefke}, {Schmitt}, {Robrade}
  \& {Ness}}{{G{\"u}nther} et~al.}{2006}]{Gunther06}
{G{\"u}nther} H.~M.,  {Liefke} C.,  {Schmitt} J.~H.~M.~M.,  {Robrade} J.,
  {Ness} J.,  2006, \aap, 459, L29

\bibitem[\protect\citeauthoryear{{Hussain}, {Collier Cameron}, {Jardine},
  {Dunstone}, {Velez}, {Stempels}, {Donati}, {Semel}, {Aulanier}, {Harries},
  {Bouvier}, {Dougados}, {Ferreira}, {Carter} \& {Lawson}}{{Hussain}
  et~al.}{2009}]{Hussain09}
{Hussain} G.~A.~J.,  {Collier Cameron} A.,  {Jardine} M.~M.,  {Dunstone} N.,
  {Velez} J.~R.,  {Stempels} H.~C.,  {Donati} J.-F.,  {Semel} M.,  {Aulanier}
  G.,  {Harries} T.,  {Bouvier} J.,  {Dougados} C.,  {Ferreira} J.,  {Carter}
  B.~D.,    {Lawson} W.~A.,  2009, \mnras, p.~997

\bibitem[\protect\citeauthoryear{{Hut}}{{Hut}}{1981}]{Hut81}
{Hut} P.,  1981, \aap, 99, 126

\bibitem[\protect\citeauthoryear{{Johns-Krull}}{{Johns-Krull}}{2007}]{Johns07}
{Johns-Krull} C.~M.,  2007, \apj, 664, 975

\bibitem[\protect\citeauthoryear{{Kaigorodov}, {Bisikalo}, {Fateeva} \&
  {Sytov}}{{Kaigorodov} et~al.}{2010}]{kaygorodov10}
{Kaigorodov} P.~V.,  {Bisikalo} D.~V.,  {Fateeva} A.~M.,    {Sytov} A.~Y.,
  2010, Astronomy Reports, 54, 1078

\bibitem[\protect\citeauthoryear{{Kurucz}}{{Kurucz}}{1993}]{Kurucz93}
{Kurucz} R.,  1993, CDROM \#~13 (ATLAS9 atmospheric models) and \#~18 (ATLAS9
  and SYNTHE routines, spectral line database).
Smithsonian Astrophysical Observatory, Washington D.C.

\bibitem[\protect\citeauthoryear{{Lanza}}{{Lanza}}{2006}]{Lanza06}
{Lanza} A.~F.,  2006, \mnras, 369, 1773

\bibitem[\protect\citeauthoryear{{Mentuch}, {Brandeker}, {van Kerkwijk},
  {Jayawardhana} \& {Hauschildt}}{{Mentuch} et~al.}{2008}]{Mentuch08}
{Mentuch} E.,  {Brandeker} A.,  {van Kerkwijk} M.~H.,  {Jayawardhana} R.,
  {Hauschildt} P.~H.,  2008, \apj, 689, 1127

\bibitem[\protect\citeauthoryear{{Mohanty}, {Jayawardhana} \&
  {Basri}}{{Mohanty} et~al.}{2005}]{Mohanty05}
{Mohanty} S.,  {Jayawardhana} R.,    {Basri} G.,  2005, \apj, 626, 498

\bibitem[\protect\citeauthoryear{{Natta}, {Testi}, {Muzerolle}, {Randich},
  {Comer{\'o}n} \& {Persi}}{{Natta} et~al.}{2004}]{Natta04}
{Natta} A.,  {Testi} L.,  {Muzerolle} J.,  {Randich} S.,  {Comer{\'o}n} F.,
  {Persi} P.,  2004, \aap, 424, 603

\bibitem[\protect\citeauthoryear{{Quast}, {Torres}, {de La Reza}, {da Silva} \&
  {Mayor}}{{Quast} et~al.}{2000}]{Quast00}
{Quast} G.~R.,  {Torres} C.~A.~O.,  {de La Reza} R.,  {da Silva} L.,    {Mayor}
  M.,  2000, in IAU Symposium Vol.~200 of IAU Symposium, {V4046 Sgr, a key
  young binary system.}.
p.~28P

\bibitem[\protect\citeauthoryear{{Rodriguez}, {Kastner}, {Wilner} \&
  {Qi}}{{Rodriguez} et~al.}{2010}]{Rodriguez10}
{Rodriguez} D.~R.,  {Kastner} J.~H.,  {Wilner} D.,    {Qi} C.,  2010, \apj,
  720, 1684

\bibitem[\protect\citeauthoryear{{Siess}, {Dufour} \& {Forestini}}{{Siess}
  et~al.}{2000}]{Siess00}
{Siess} L.,  {Dufour} E.,    {Forestini} M.,  2000, \aap, 358, 593

\bibitem[\protect\citeauthoryear{{Stempels} \& {Gahm}}{{Stempels} \&
  {Gahm}}{2004}]{Stempels04}
{Stempels} H.~C.,  {Gahm} G.~F.,  2004, \aap, 421, 1159

\bibitem[\protect\citeauthoryear{{Torres}, {Quast}, {Melo} \&
  {Sterzik}}{{Torres} et~al.}{2008}]{Torres08}
{Torres} C.~A.~O.,  {Quast} G.~R.,  {Melo} C.~H.~F.,    {Sterzik} M.~F.,  2008,
  {Young Nearby Loose Associations}.
p.~757

\bibitem[\protect\citeauthoryear{{Yee} \& {Jensen}}{{Yee} \&
  {Jensen}}{2010}]{Yee10}
{Yee} J.~C.,  {Jensen} E.~L.~N.,  2010, \apj, 711, 303

\bibitem[\protect\citeauthoryear{{Zahn} \& {Bouchet}}{{Zahn} \&
  {Bouchet}}{1989}]{Zahn89}
{Zahn} J.-P.,  {Bouchet} L.,  1989, \aap, 223, 112

\end{thebibliography}
\bibliographystyle{mn2e}
\end{document}